\documentclass[sn-mathphys]{sn-jnl}% Math and Physical Sciences Reference Style
\jyear{2021}%

\theoremstyle{thmstyleone}%
\theoremstyle{thmstyletwo}%

\theoremstyle{thmstylethree}%
\newcommand{\m}[1]{#1\,\text{m}}

\newcommand{\mm}[1]{#1\,\text{mm}}
\newcommand{\s}[1]{#1\,\text{s}}
\newcommand{\ms}[1]{#1\,\text{ms}\textsuperscript{-1}}

\begin{document}

\title[Article Title]{Experimental Investigation of Droplet Generation by Post-Breaking Plunger Waves}


\author*[1]{\fnm{R.G. Ramirez de la Torre}} \email{reynar@math.uio.no}

\author[2]{\fnm{Petter Vollestad}} \email{pettervollestad@hotmail.com}
\equalcont{These authors contributed equally to this work.}

\author[1]{\fnm{Atle Jensen}} \email{atlej@math.uio.no}
\equalcont{These authors contributed equally to this work.}

\affil*[1]{\orgdiv{Department of Mathematics}, \orgname{University of Oslo}, \country{Norway}}
\affil*[2]{\orgname{Det Norske Veritas}, \country{Norway}}


\abstract{Understanding the droplet cloud and spray dynamics is important for the study of the ocean surface and marine boundary layer. The role that the wave energy and the type of wave breaking play in the resulting distribution and dynamics of droplets is yet to be understood. The aim of this work was to generate violent plunging breakers in the laboratory and analyze the spray production post-breaking, i.e. after the crest of the wave impacts in the free surface. The droplet sizes and their dynamics were measured with imaging techniques and the effect of different wind speeds on the droplet production was also considered. It was found that the mean radius increases with the wave energy content and the number of larger droplets \textcolor{black}{(radius $>$ \mm{1})} in the vertical direction increases with the presence of wind. Furthermore, the normalized distribution of droplet sizes is consistent with the distribution of ligament-mediated spray formation. \textcolor{black}{Also, indications of turbulence affecting the droplet dynamics at wind speeds of \ms{5} were found. The amount of large droplets (radius $>$ \mm{1}) found in this work was larger than reported in field studies.} %However, the analyzed systems in field studies are different than the presented in this work.
}




\maketitle

\section{Introduction}\label{sec1}
At the ocean surface, a variety of complex two phase flow interactions generate aeration inside the water and aerosol transport through the air. In the present study, we are interested in wave breaking related to marine icing processes. For example, in the Arctic environment the droplets produced after wave breaking are transported by the wind and generate thick layers of ice over the surface of ships and structures. These ice layers represent a life hazard for the inhabitants of these vessels. 
Field studies and simulations have been used to address this phenomenon \citep{revMarineIcing,dehghani2016vel,bodaghkhani2016understanding,ryerson1990atmospheric}, \textcolor{black}{it is clear that the main source of marine icing is breaking waves, but a deeper understanding of the droplet generation post-breaking is necessary.} The study of droplet size distribution and dynamics is also important to understand the transport of heat and momentum through the marine boundary layer and above. To model this phenomenon, detailed information about the size and velocity distributions of the droplets is needed.  
Small droplets can be transported over long distances and remain in the atmosphere for several days, while large droplets remain close to the ocean surface and return to the ocean on shorter time scales but droplets of all size affect the air-sea fluxes of momentum and enthalpy \citep{veron2015ocean}.

In the review of Ocean Spray \cite{veron2015ocean} previous findings and emerging consensus on sea spray generation were summarized, generation processes for droplets with radius up to \mm{1} were thoroughly analyzed. These small droplets (radius $<$ \mm{1}) have residence times in the atmosphere from minutes to several days, or even weeks --when the radius is only around 10 nm. The long residence times allow to make direct estimations of the spray size dependence on the wind velocities by measuring the drop concentration average profile through time. The review also summarizes thoroughly the studies over direct and indirect methods to estimate a Sea Spray Generation Function (SSGF). It is pointed out, that indirect estimations of the SSGF have the common assumption of a universal source function of droplet sizes, and that the change on number density for a particular size range is considered to depend only on other controlling parameters, such as wind speed, fetch, surface stress, etc. Other studies have confirmed that a source function can be used to estimate the shape of the SSGF \cite{villermaux2004ligament,mueller2009sea}. The review closes by highlighting that one of the main issues to study in the future is the production of large spume droplets (radius larger than \mm{1}), their generation mechanism, initial velocity and dynamic behaviour through the airflow. Moreover, field studies of droplet distribution on vessels showed that the size distributions extend to several millimeters \citep{bodaghkhani2016understanding,ryerson1990atmospheric}. 
The present study is an attempt to contribute to the understanding of the large droplet behaviour. In particular the generation mechanism, initial size distribution and the dynamic behaviour in the airflow.

There are several studies of droplet size distribution available, in particular, there have been studies where the importance of the initial distribution or source function was addressed. \textcolor{black}{In one of these studies, a $\Gamma$-distribution was proposed to fit the droplet size distribution created after the break-up and coalescence of so called \textit{ligaments} that detached from the main water bulk of a round jet \cite{villermaux2004ligament}.} The dependence of the droplet distribution on the volume and diameter of these ligaments independently of the shape of the liquid bulk was presented. Then, the proposed $\Gamma$-distribution was used as the source function to calculate the shape of the SSFG \cite{mueller2009sea}. The proposed function implied considerably large energy fluxes at low and moderate winds. These findings remark the importance of the individual processes of generation and suspension of droplets and underlines the complexity of the initial size distributions due to the variety of generation processes.
 
More recently, an experimental study with mechanical waves and winds up to \ms{54}was developed \cite{ortiz2016sea}. The findings showed that for droplets with radii $\sim$ \mm{1}, the production rates were several orders of magnitude higher than the rates expected from previous investigations \citep{fairall2009investigation,veron2015ocean}. \textcolor{black}{The droplets were measured at locations between 2 and 6 times the local significant wave height, and for the highest wind speeds droplets with radius $\sim$ \mm{1} were observed in relatively high quantities at heights between 3 and 4 times the significant wave height.} Furthermore, field measurements have been conducted, where the concentration of aerosol numbers in the atmospheric boundary layer were obtained \citep{lenain2017evidence}, droplet sizes ranging from 0.1 to 200 microns were measured. It was found that droplets with radii larger than 40 microns can reach heights higher than 400 m above mean sea level. These findings may suggest that large droplets have a longer lifetime in the atmospheric boundary layer than previously expected. Therefore, the processes that generate larger droplets and allow the relatively long lifetimes need to be better understood.

The importance of the dynamics of the droplet generation and transport has also been studied. The description of dispersion and transport of droplets has been done by examining the motion of a single drop and quantifying the influence of the airflow and turbulence over the droplet. Equations for terminal velocities and drag coefficients have been obtained and related to particle Reynolds numbers ($Re$), Stokes numbers ($St$) and the Kolmogorov time scaling \citep{clift2005bubbles,andreas2010production,crowe2011multiphase}. But due to the large number of droplets that can be produced in one event, it is also important to consider the statistics of the phenomena. In general, particles moving in a fully developed turbulent flow have velocity components that are normally distributed and the speed follows the Maxwell-Boltzman (\textit{M-B}) distribution, similarly to the Brownian motion \citep{pope1994lagrangian}. Also, it has been found that the acceleration components has a stretched exponential shape with largely extended tails compared to a normal distribution \citep{la2001fluid}. \textcolor{black}{This is a phenomenological function for particles travelling in flows with $200 \leq R_\lambda \leq 970$, where as $R_\lambda = (15Re_{f})^{1/2}$ is the Taylor microscale Reynolds for the turbulent length scale $\lambda$, defined in terms of the Reynolds number of the flow that surrounds the particles $Re_{f} = LU/\nu$, where $L$ is the characteristic length, $U$ is the velocity of the flow and $\nu$ is the kinematic viscosity.} This function has been experimentally confirmed by different studies in various fluid dynamics applications \citep{voth2001measurement,mordant2004experimental,shnapp2019extended,kim2019lagrangian}. 

\textcolor{black}{In this study we present experimental results for medium ($\mm{0.25} \leq r \leq \mm{1}$) and large droplets ($ \mm{1} \leq r  \leq \mm{5.5}$) generated by plunging breakers.} When the crest of the plunging breakers impact the free surface, a large quantity of spray is produced. Cases without wind and with the winds between 3 and \ms{7} have been studied. The aim is to identify the shape of the initial size distribution, or source function, and relate the conditions at the source (like wave energy content and wind) to the dynamics of the droplets. Our work is structured as follows. In section 2, the experimental setup is presented. \textcolor{black}{In section 3, we present results and the discussion; the first part presents the size distributions obtained by means of a video/image analysis method developed in house, and in the second part the droplet dynamics obtained by means of three dimensional Particle Tracking Velocimetry (3D PTV) is discussed. These results are analyzed and discussed to obtain statistical distributions of initial droplet diameter, vertical reach, velocity and accelerations. Finally, section 4 presents the conclusion of this work.}

\section{Experimental setup}\label{sec2}
\begin{figure}
	\includegraphics[width=\textwidth]{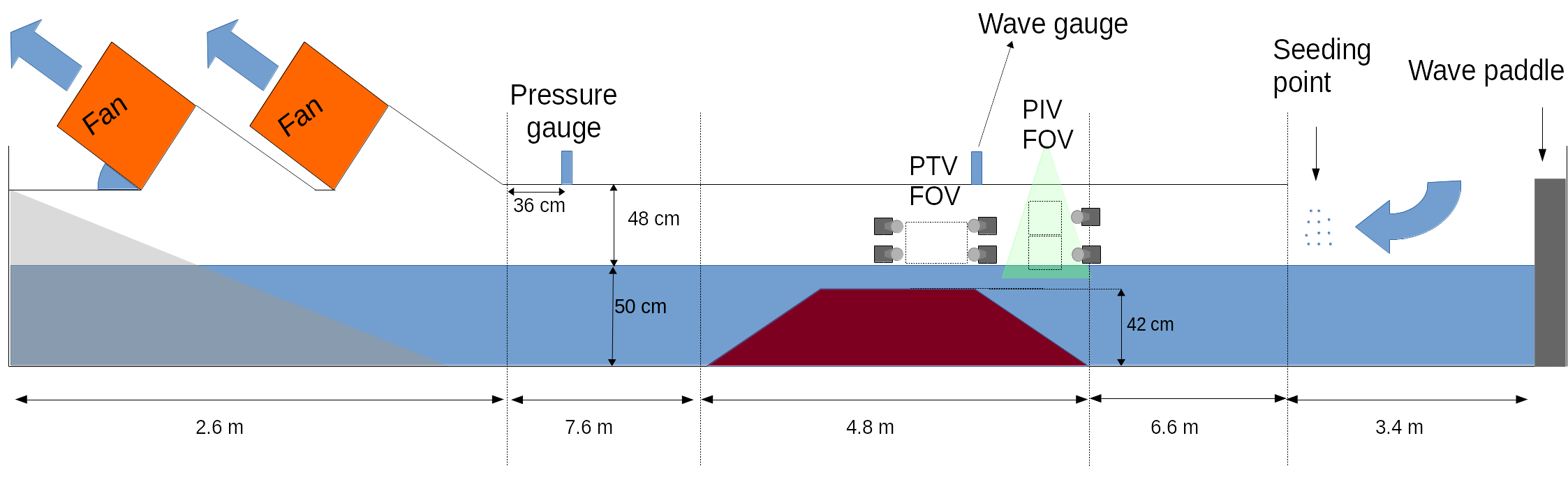}
	\caption{\label{fig:ExpSetup}Schematic drawing of the wave tank in the Hydrodynamics Laboratory where the experiments were developed}
\end{figure}
The experiments were conducted in the Hydrodynamics Laboratory at the University of Oslo, in a wave tank with dimensions $25\times0.52\times\m{1}$ where the mean water level for all experiments was $\m{0.5}$, as shown in Figure \ref{fig:ExpSetup}. In this work, violent plunging breakers are created and the produced droplets post-breaking are quantified, obtaining their sizes and dynamics and analyzing the effect of the different wind speeds on the droplet production. To produce breaking waves, a focusing wave train is used, where long waves overtake short waves as proposed by \cite{brown2001experiments}. The wave breaking process was made violent by adding a slope and shoal area to the bathymetry which caused the already focused waves to steepen and overturn. After breaking, the overturning crest of the wave splashed at the free surface releasing a large number of droplets. A detailed description of the wave generation can be found in Appendix \ref{sec:A-wave} and the effects of the shoal area can be found in Appendix \ref{sec:A-shoal}. The wind velocity profiles produced on top of the waves and its influence in the wave train are also described in the Appendix \ref{sec:A-wind} and \ref{sec:A-ww}. 

\begin{figure}
	\includegraphics[width=0.5\textwidth]{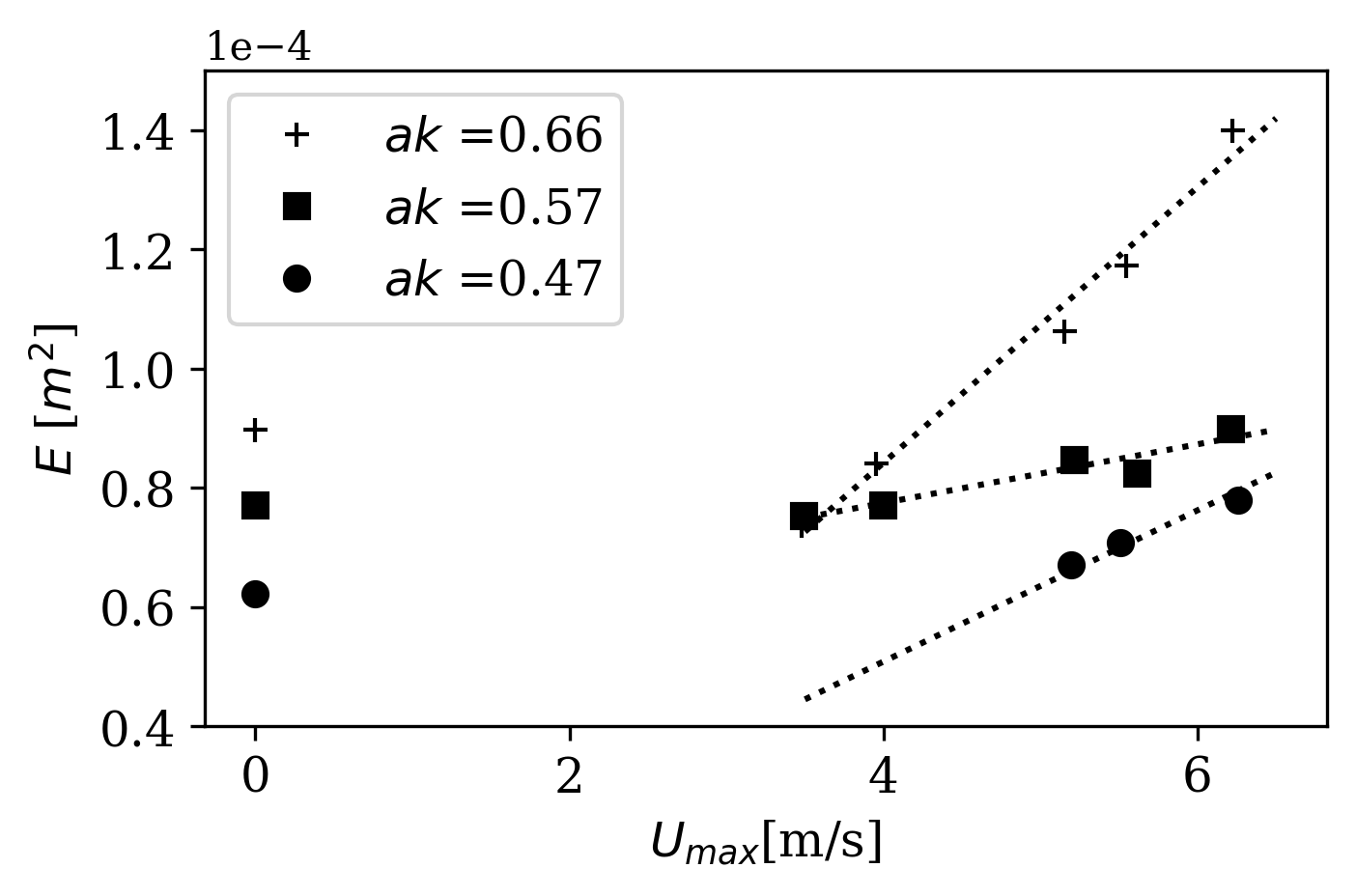}
	\caption{Mean power of the wave series against wind speed. The different markers represents different wave steepness $ak$.}
	\label{fig:all_waves}
\end{figure}

The experiments developed as follows: a focusing wave is generated and focused at the edge of the slope; when the focused packet approaches the shoal area, the slope and wind presence affects the energy content of the wave group (see, details in Appendix \ref{sec:A-ww}). As a result, the wave steepens faster overturning and splashing in the shoal area, where droplets are generated. These droplets are recorded by a 3D PTV 4-camera system. The energy of the breaker is quantified and compared to the droplet production. \textcolor{black}{To quantify the change in energy content of the different cases, we can use the mean power, $E$, as defined by statistics \citep{holthuijsen2010waves}:
\begin{equation}
	E = \int_{-\infty}^{\infty} S(f)df,
\end{equation}
where $E$ is the area under the spectral curve $S(f)$, for the different wave frequencies $f$. $E$ can be interpreted as the energy content of the wave, as $E$ is proportional to the wave amplitude squared ($a^2$) which is also proportional to the energy. Figure \ref{fig:all_waves} shows the calculated $E$ compared to the wind velocities:} $U_{max}$, and different maximum wave steepness $ak$, where $k$ is the wave number calculated for the dispersion relation (see Appendix \ref{sec:A-wave}). The graph shows the effect of wind over the wave energy. In all cases the energy increases with $ak$. But, it is interesting to see that for $U_{max} < \ms{4.5}$ the total energy of the packet is less than the energy of the packet without the presence of wind. 
\textcolor{black}{It is also interesting to notice the different gradient for the cases with $ak=0.57$, the mean power or energy content of these cases seems to vary at a lower rate than the other cases. We believe that this might be a result of the underlying non linear effects of the wave packet and its interaction with the slope. Nonetheless, these effects require further study that will not be included in this work.} 
Table \ref{tab:break} shows the breaking type for all the cases analyzed. The difference between a small plunger and a plunger is the plunge distance. The plunge distance is defined as the distance from the break point to the crest touchdown point. We call the breaking type ``small plunger" if the plunge distance is smaller than $a_{max}/2$. The case of $ak=0.47$ with wind velocities $U_{max} < \ms{4.5}$  do not generate a plunging breaker, therefore this data is not included in the study.

\begin{table}
	\caption{Type of breaking before and after adding the slope and under different wind conditions.}\label{tab:break}
	\centering
	\begin{tabular}{l c c c}
		\hline
		Conditions  & $ak = 0.47$ & $ak=0.57$ & $ak=0.66$ \\
		\hline
		no shoal  & spilling & spilling &spilling\\
		\hline
		shoal, $U_{max}=0$     & spilling & small plunger & small plunger\\
		shoal, $U_{max}=3.41$  & spilling & small plunger &plunger\\
		shoal, $U_{max}=3.91$  & spilling & plunger &plunger\\
		shoal, $U_{max}=5.14$  & small plunger & plunger &plunger\\
		shoal, $U_{max}=5.45$  & plunger & plunger &plunger\\
		shoal, $U_{max}=6.22$  & plunger & plunger &plunger\\ 
		\hline
	\end{tabular}
\end{table}

\subsection{3D PTV}
\begin{figure}
	\includegraphics[width=0.65\textwidth]{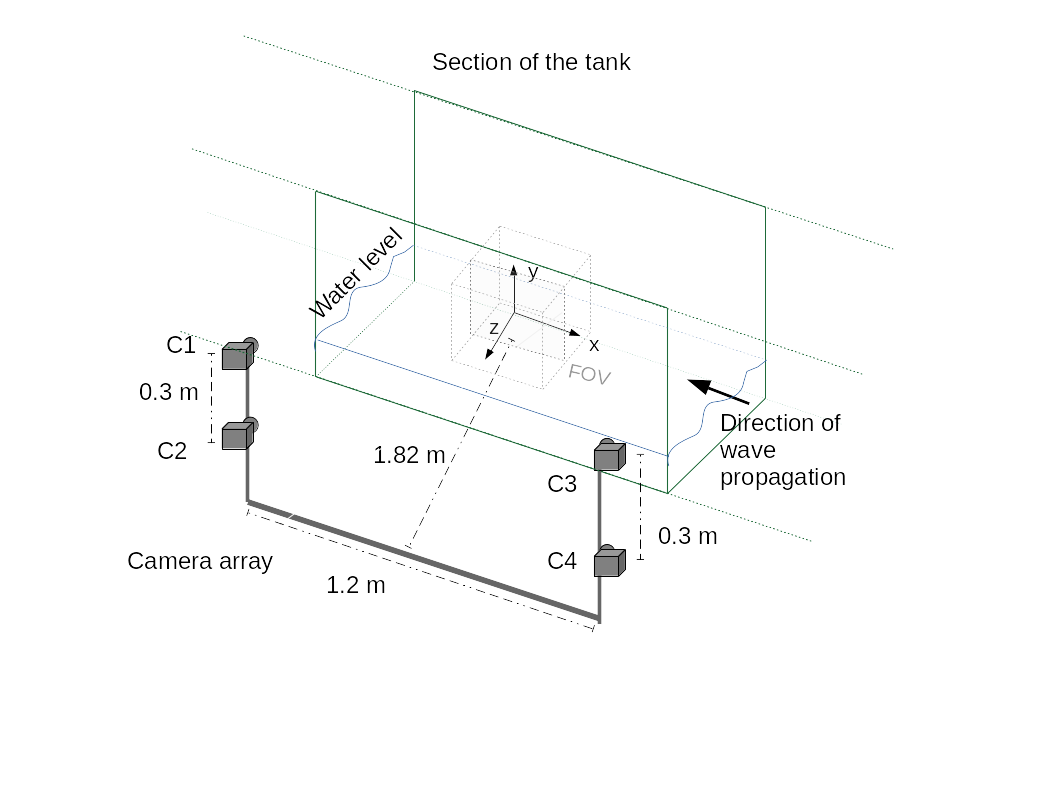}
	\hspace*{-0.4in}
	\raisebox{0.5\height}{\includegraphics[width=0.35\textwidth]{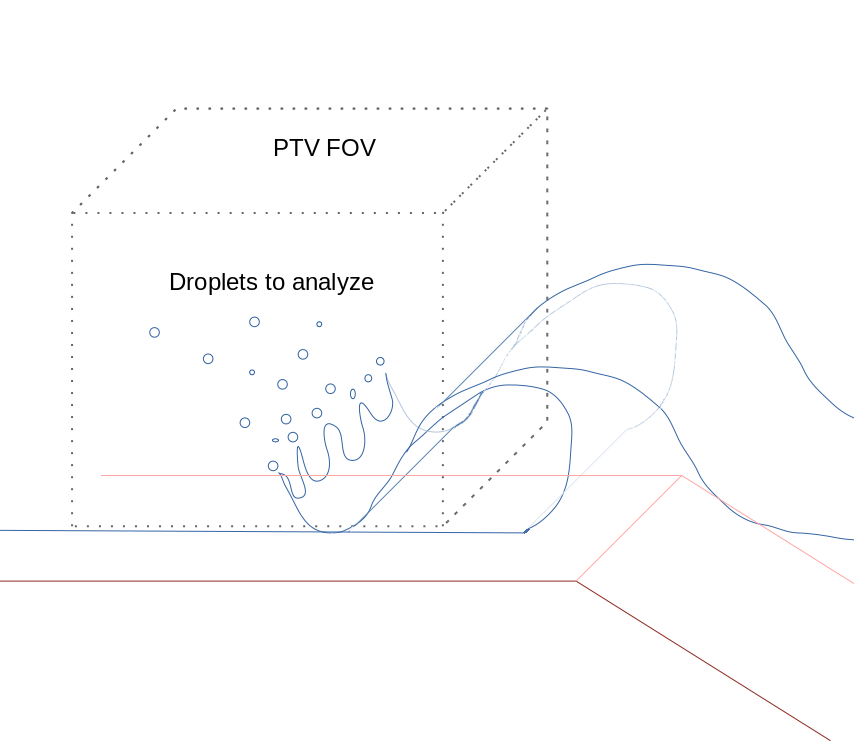}}
	\caption{To the left, schematic drawing of the 4-camera setup for 3D PTV, the Field of View for PTV is shown by the line-delimited volume, the direction of the coordinate system is depicted. It is important to notice that wave propagates towards negative $x$. To the right, a schematic example of the experiments and the field of view where the droplets are captured by the cameras.}
	\label{fig:ptv}
\end{figure}
%When the wave breaks, the crest accelerates and curls over the front. Then the crest impacts the free surface in the shoal area and splashes, creating a large number of droplets. 
After the wave breaks, the disintegrating wave keeps moving forward and ejecting droplets from its surface. We analyze only visible droplets in the selected FOV as shown in Figure \ref{fig:ptv}, mainly the droplets generated by the splashing crest. The trajectories, velocities and accelerations of the droplets are obtained using 3D PTV. \textcolor{black}{The particle tracking algorithm follows the trajectories of each particle, or in this case droplets, through space and time in the Lagrangian framework.} Coordinates in three dimensions can be obtained by using stereoscopic imaging and synchronous recording of the motion. The droplet positions are tracked in the time domain to derive the velocities and accelerations as differences in position between consecutive time steps. To estimate the size of the droplets, a combined method that uses the information from PTV results and the images was developed in house \cite{3dptv-sizes}. The method consists of making an average of the pixel size of each droplet in the images obtained by the four cameras, then correlating the pixel size to the position of the droplet determined by PTV, in this way a relation between the pixel size and the size in millimeters can be obtained under certain assumptions. The obtained size estimations have a maximum error of 10 \% in all cases.

A 4-camera system is used to perform 3D PTV, using the open source software OpenPTV \citep{openptv2012openptv}. Images of the droplets are taken by 4 Monochromatic AOS Promon cameras with 50mm lenses. The frame rate is 167 fps and the image resolution is $1920 \times 1080$ pixels with a resolution of \mm{0.15} per pixel. With this resolution, the smallest droplets that can be detected have a diameter of \mm{0.45} (equivalent to 3 pixels) \textcolor{black}{and velocities between \ms{0.075} and \ms{12} can be observed}. The FOV right side is located on the breaking point to obtain all the splashing occurred in front of the wave, as shown in Figure \ref{fig:ExpSetup}. The three-dimensional FOV is approximately $0.25 \times 0.15 \times \m{0.20} $, as shown in Figure \ref{fig:ptv} where the FOV is represented by the gray outlined area. The gray plane represents the focal plane of the cameras which corresponds to $z = 0$. It is important to notice the direction of the waves and the wind, which is in the negative direction of the x-axis. \textcolor{black}{A sequence of 2 seconds from the moment that the wave reaches the focal point  is recorded, this cover the whole duration of the wave breaking process in all cases.} From the post-processing we can also obtain size distributions of the droplet cloud. A set of 5 repetitions was developed for each wave amplitude and wind speed. 

\section{Results and Discussion}\label{sec3}
The splashing and spray generation process happens in a span of around one second, and the physical event has an inherent randomness. Therefore the results of the 5 experiments are used as a statistical ensemble. For each droplet we collect the results of the PTV processing (size, position, velocity and acceleration) in each time step. Every time step, from the plunge to the collapse of the wave, is considered in the analysis. The droplets are not always spherical and their deformation increases with the size. \textcolor{black}{Therefore, the equivalent diameter $D_e$ is commonly used to classify droplet sizes with one unique parameter and is commonly defined as $D_e = \sqrt{l_{M} l_{m}}$, 
where $l_{M}$ and $l_{m}$ are the major and minor axis of the ellipse, in this case $l_{M}$ and $l_{m}$ are estimated from the 4 images of the camera array and an average of these 4 images is obtained.} 

In this section, the relations between the wave energy, the wind velocity and the production of droplets will be presented. First, the size distribution of the droplets will be related to the wave energy and wind. Then similar relations will be shown for the velocity and the acceleration of the droplets in the different cases of wave energy and wind speed.

\subsection{Droplet Sizes}
\begin{figure}[ht]%  figure placement: here, top, bottom, or page
	\centering
	\begin{picture}(100,50)
		\put(60,25){No wind}
	\end{picture}
	\begin{picture}(100,50)
		\put(30,25){$U_{max}=\ms{5.2}$}
	\end{picture}
	\begin{picture}(100,50)
		\put(30,25){$U_{max}=\ms{6.2}$}
	\end{picture}
	\vspace{-10.0mm} 
	
	\begin{picture}(30,100)
		\put(0,10){\rotatebox{90}{ak = 0.47}}
	\end{picture}
	\includegraphics[width=0.25\textwidth]{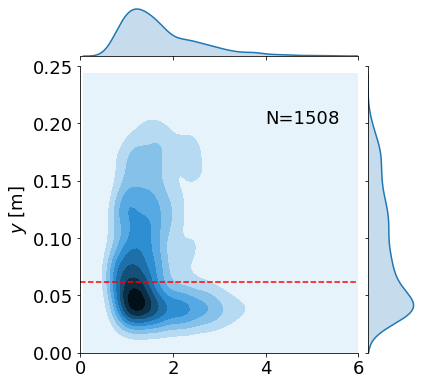}
	\includegraphics[width=0.25\textwidth]{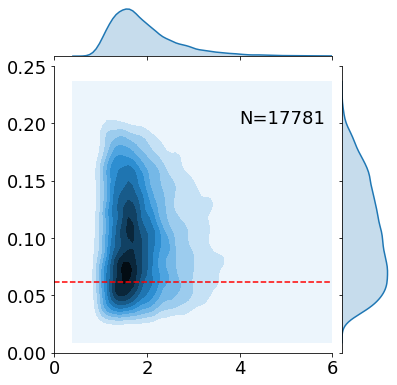}
	\includegraphics[width=0.25\textwidth]{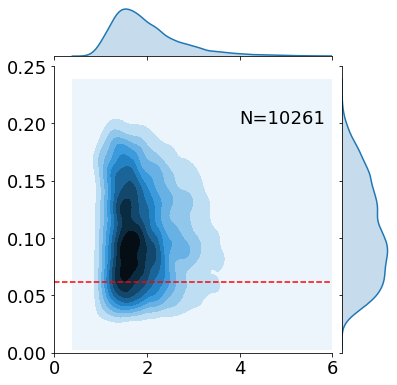}   \\
	\begin{picture}(30,100)
		\put(0,10){\rotatebox{90}{ak = 0.57}}
	\end{picture}
	\includegraphics[width=0.25\textwidth]{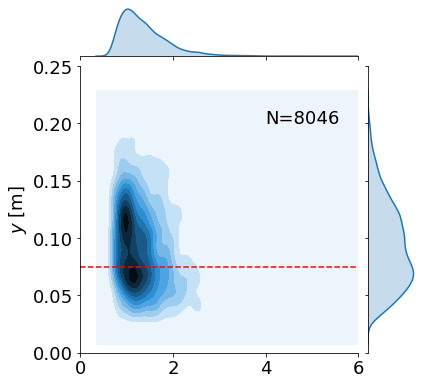} 
	\includegraphics[width=0.25\textwidth]{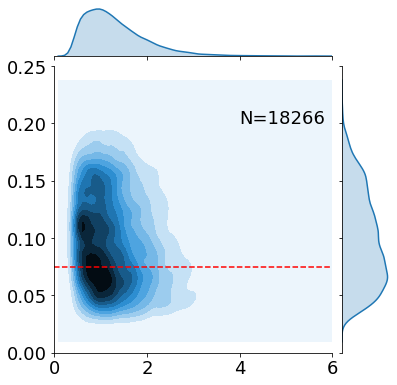}
	\includegraphics[width=0.25\textwidth]{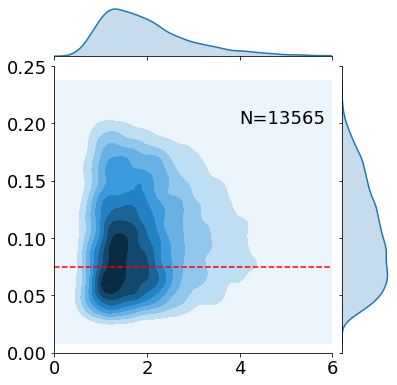}   \\
	\begin{picture}(30,100)
		\put(0,10){\rotatebox{90}{ak = 0.66}}
	\end{picture}
	\includegraphics[width=0.25\textwidth]{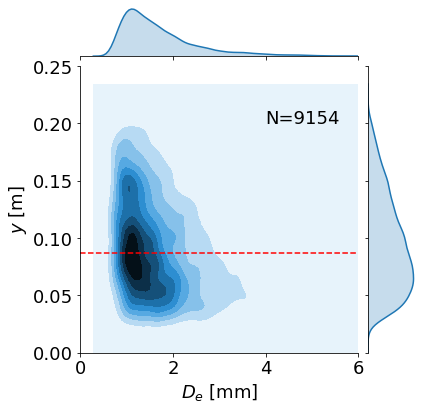} 
	\includegraphics[width=0.25\textwidth]{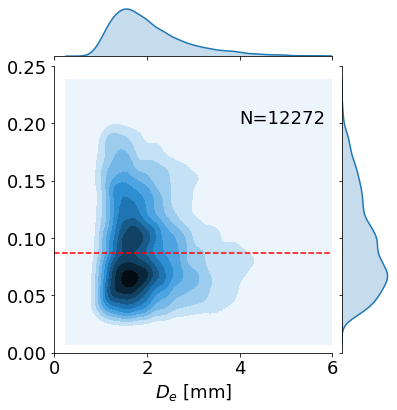}
	\includegraphics[width=0.25\textwidth]{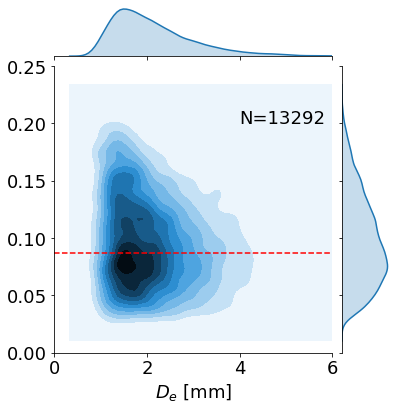}  \\
	\includegraphics[width=\textwidth]{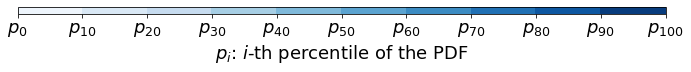}
	
	\caption{Bi-variate probability distribution of droplets with Equivalent Diameter $D_e$ and height $y$ as variables. In the margins, the uni-variate, or projected, distribution of droplets are shown: the PDF by size or $D_e$ is presented on the top and on the side the PDF by height is presented. The rows show different steepness $ak$ and the columns show wind velocities $U_{max}$. The red dotted line represents the maximum wave height before breaking. The colors represents different contour levels, or $i$-th percentiles, where every level between $p_i$ and $p_{i+10}$ contains 10\% of the total detected droplets.}
	\label{fig:Sizes}
\end{figure}
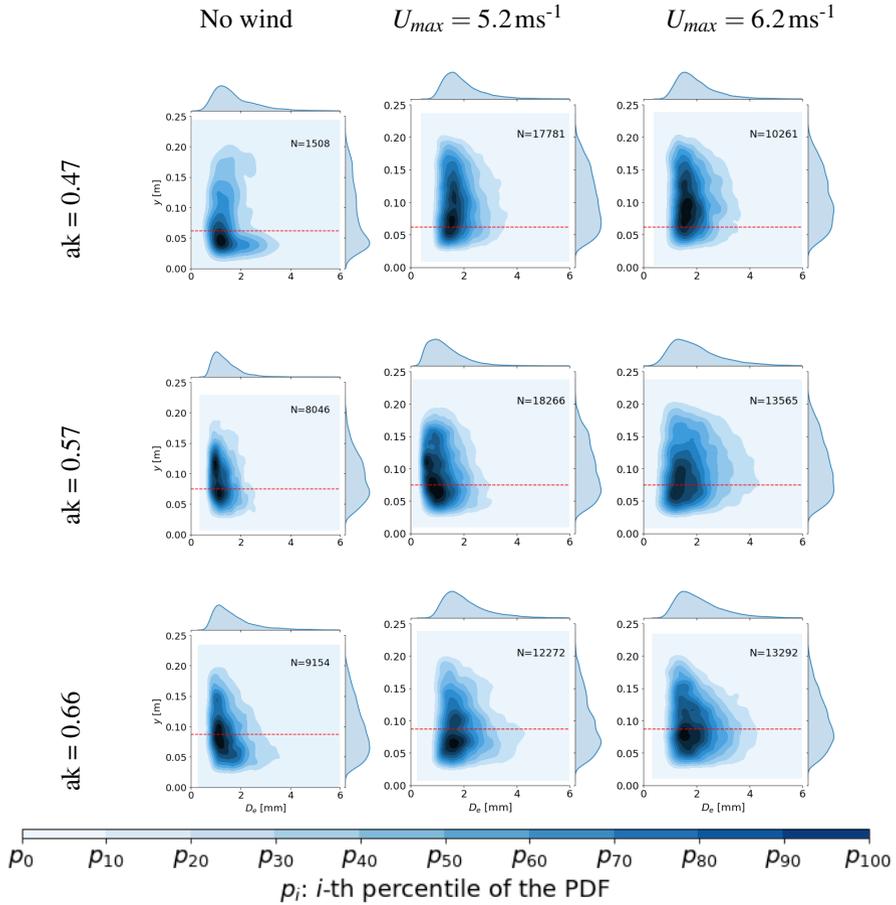

\begin{figure}
	\includegraphics[width=\textwidth]{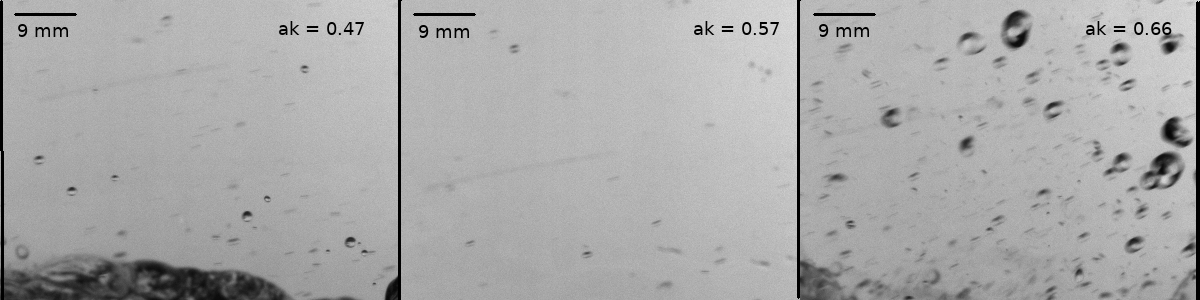}
	\caption{\textcolor{black}{Example of droplet image captured for the different cases. Similar FOV on top of the waves is shown for the three images. It is easy to see, that the case $ak=0.57$ produces droplets with a size comparable to that of $ak=0.47$ and visibly smaller thn $ak=0.66$.}}
	\label{fig:drops}
\end{figure}

Figure \ref{fig:Sizes} shows the equivalent diameter $D_e$ and height distributions of droplets for different cases of $ak$ and $U_{max}$. In all cases, higher concentrations of larger droplets are presented when the wind is applied. \textcolor{black}{When $U_{max}=0$ the droplets with $D_e > \mm{2}$ are clearly found only under $y \leq a_{max}$. In contrast, larger concentrations of these droplets are found over $y = a_{max}$ for the wind cases.} This result agrees with the hypotheses that more energetic waves will produce droplets with larger speed and that more droplets will be transported further by the wind. When $ak=0.57$ the presence of large droplets is small compared to the other cases, this might be a consequence of the different types of breaking mechanisms. In this case the small plunger seems to produce less droplets than the spilling breaker. By visual inspection \textcolor{black}{(fig. \ref{fig:drops})}, the amount of spray is different, but it is difficult to quantify the difference in the breaking process.

\begin{figure}
	\includegraphics[width=0.5\textwidth]{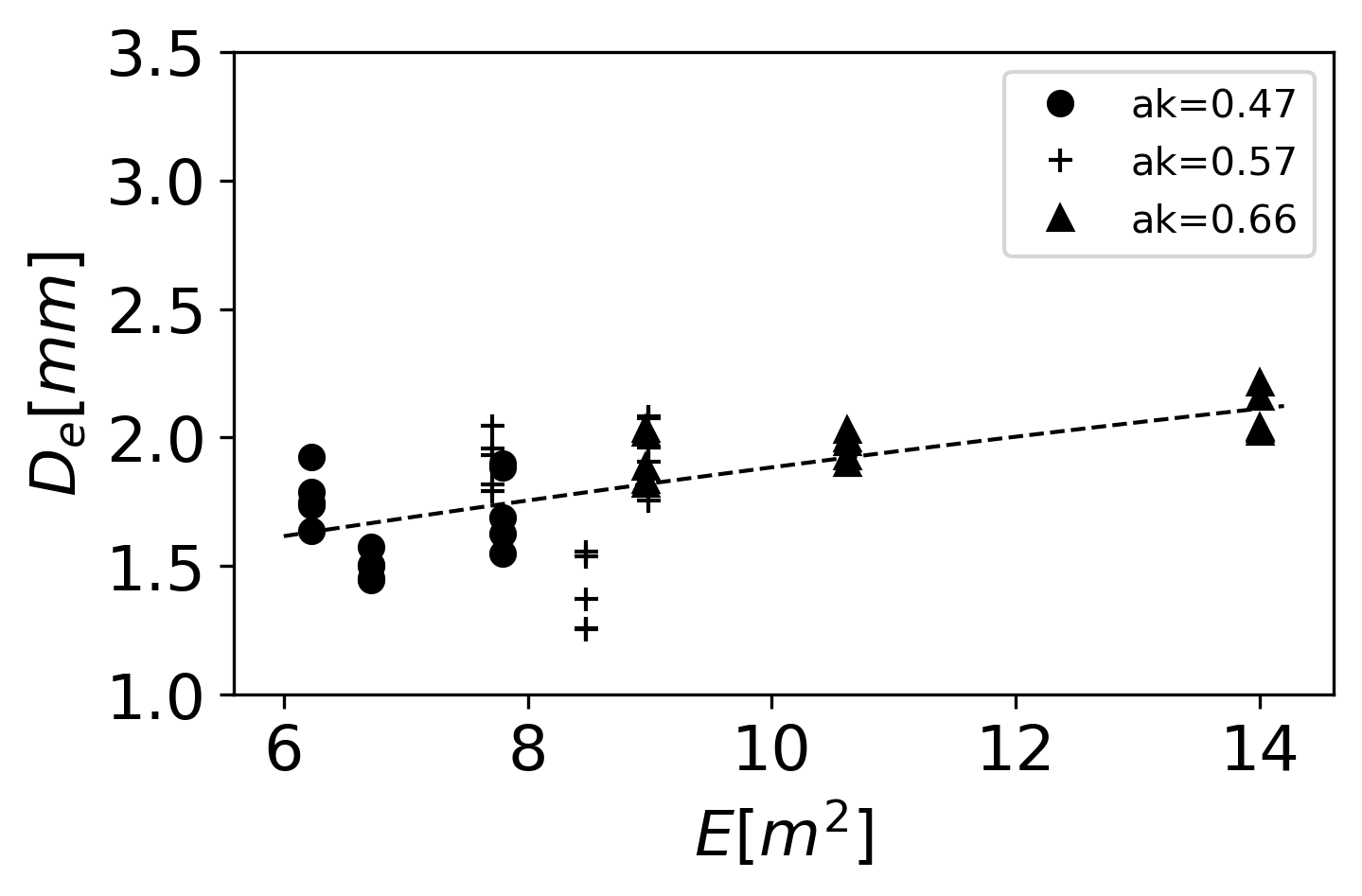}
	\caption{$E$ against mean equivalent diameter for each experiment. A linear fit is shown with the dotted line: $\overline{D_e} = 7.65 E + 0.001$, with $R^2 = 0.64$. The mean diameter seems to increase with the energy content of the wave packet.}
	\label{fig:Pow-diam}
\end{figure}

Figure \ref{fig:Pow-diam} shows $\overline{D_e}$, the mean equivalent diameter of the ensemble for the different cases in relation with energy content of the wave. $\overline{D_e}$ increases with $E$, which means that $\overline{D_e}$ depends on $ak$ and $U_{max}$. A first order polynomial fit can be made: $\overline{D_e} = 7.65 E + 0.001$ with $R^2 = 0.64$. According to \cite{mueller2009sea}: $D_0 \approx 2.5 D_e$, where $D_0$ is the diameter of a sphere with the equivalent volume as the average ligament, assuming that this relation is sustainable in the present study, then $D_0 \propto E$, which suggests that $D_0$ increases with the energy of the wave packet. Previously, it has been found that the mean size of droplets decreases with the presence of high winds \citep{mueller2009sea, ortiz2016sea,fairall2009investigation}. Our findings suggest that it is the break-up of larger droplets in the turbulent flows that contributes to the generation of smaller droplets. Therefore the study of large droplets breakup in high wind could be of interest.

\begin{figure}
	\includegraphics[width=\textwidth]{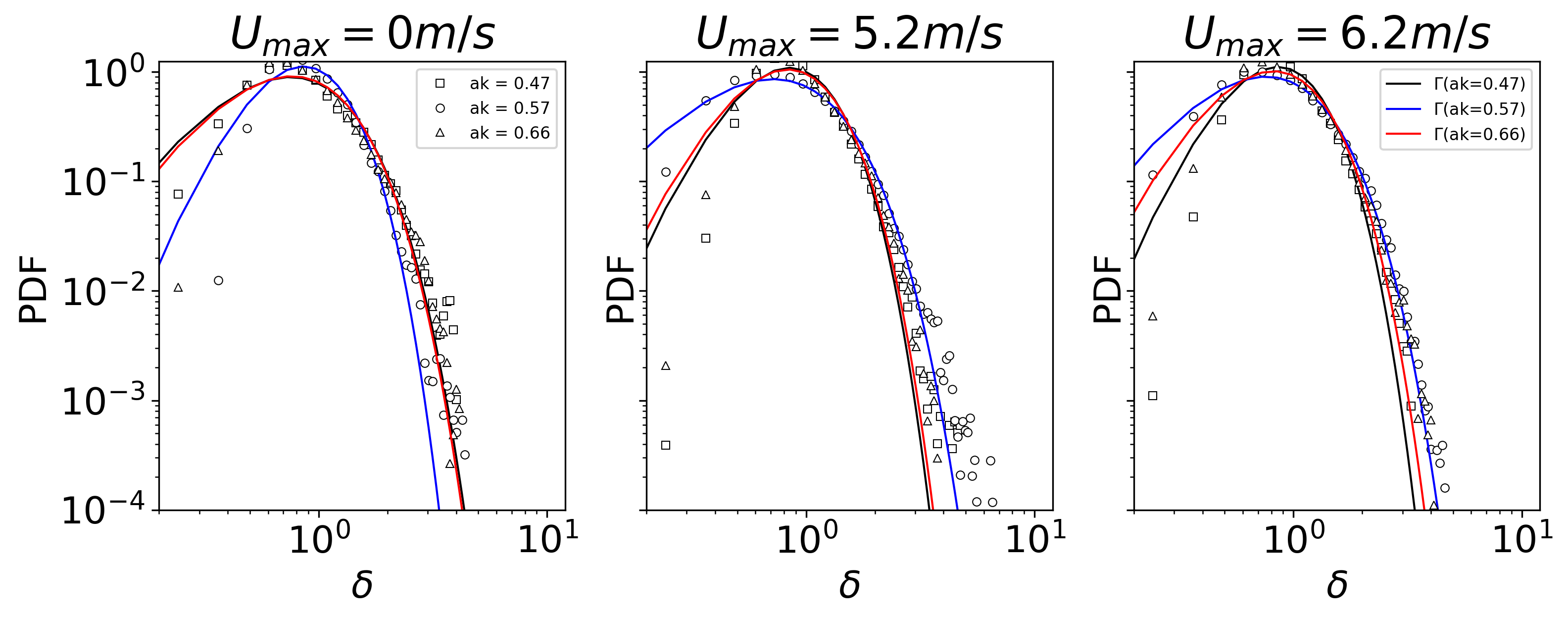}
	\caption{Probability Distribution of experimental $D_e$ compared to distribution defined in equation \ref{eq:gamma}. The data has been normalized by the mean of each experimental case so that $\delta =D_e/\overline{D_e}$. The panels shows the results for different wind speed and the marker shapes show the different steepness. The color lines shows the closest fit of equation \ref{eq:gamma} ($\Gamma$ distribution) to the data for each steepness.}
	\label{fig:size-theo}
\end{figure}
\begin{figure}[ht]%  figure placement: here, top, bottom, or page
	\centering
	\includegraphics[width=0.45\textwidth]{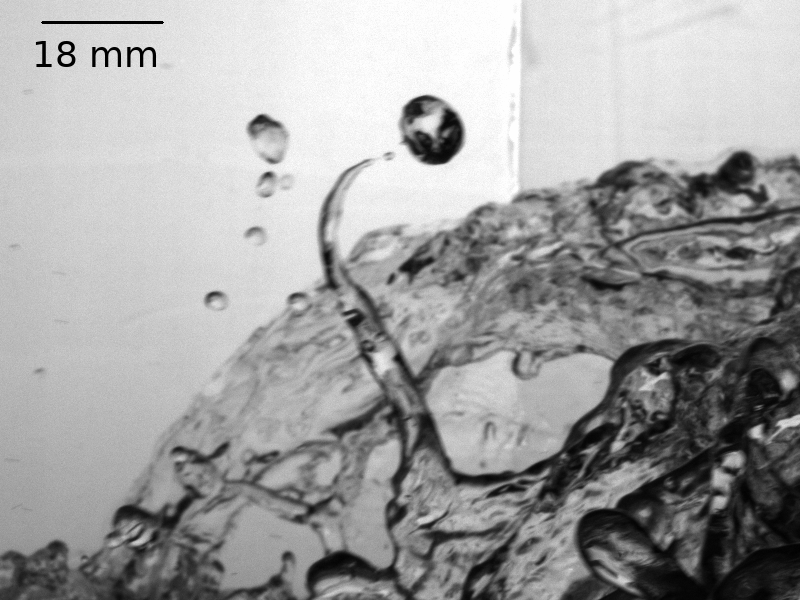}
	\includegraphics[width=0.45\textwidth]{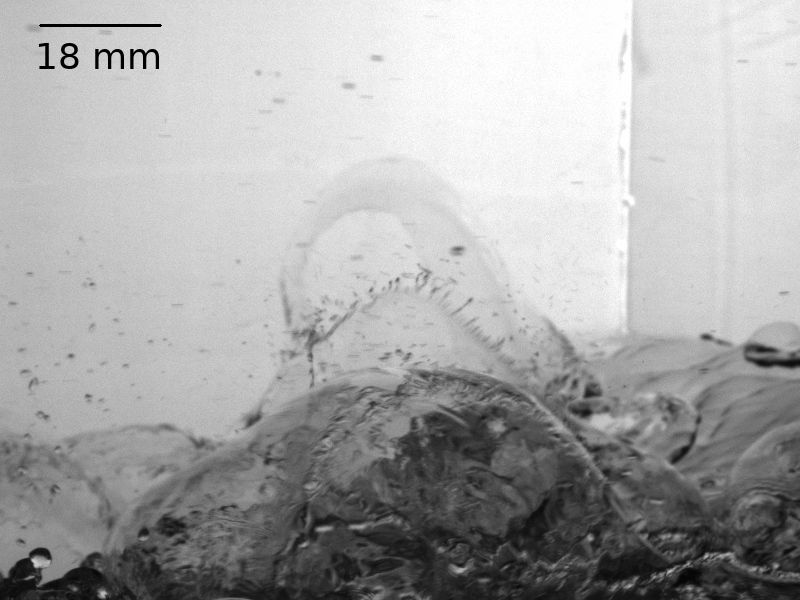}
	\caption{Examples of visible ligaments during the investigation, the first image shows a large ligament created from the wave splash, the second image shows small ligaments derived from a water film created during the collapse of the wave.}
	\label{fig:lig}
\end{figure}

\textcolor{black}{Figure \ref{fig:size-theo} shows the probability distribution of the normalized droplet diameter $\delta = D_e/\overline{D_e}$ for all cases, by normalizing the distribution all results are comparable independently of $ak$ and $U_{max}$. Comparison to other studies can also be done, the solid lines correspond to the proposed $\Gamma$-distributions \cite{villermaux2004ligament}:}
\begin{equation}\label{eq:gamma}
	\Gamma(\delta;n)=\frac{n^n \delta^{n-1} e^{-n\delta}}{\Gamma (n)}
\end{equation}
where $n^{-1}$ is the variance of $\delta$. The experimental values of $n$ obtained in this work lie between 3.5 and 7 and are similar to those in the literature  \cite{villermaux2004ligament}. A simplified relation between $n$ and the ratio $\overline{D_e}/\xi$, where $\xi$ is the average diameter of a ligament was also presented \cite{mueller2009sea}. This relation can be expressed as:
\begin{equation}\label{eq:lig}
n = 0.4(\overline{D_e}/\xi)+2
\end{equation}
In our experiments ligaments are also created during the splash and after the wave breaking, ligaments with diameters between 1--\mm{5} have been found by manual inspection of the obtained images (Fig. \ref{fig:lig}). \textcolor{black}{Although the studied system  in this work (plunging wave) is not the same as that presented in the aforementioned study(circular jet) from which ligaments are generated, the mechanism that forms droplets from the breakup of ligaments is suspected to be similar.} Assuming that most of the droplets were generated by ligament breaking, and considering equation \ref{eq:lig} holds for this study, we can assume that the droplets with mean diameter $\overline{D_e}$ come from ligaments of diameter $\xi \approx \mm{0.4}$. With a resolution of \mm{0.15} per pixel in the images acquired, most of this ligaments would be barely detectable by visual inspection in the images.
%In contrast, for this investigation, the relation has the shape: $n = 12.34(\overline{D_e}/\xi)-2$. 

\subsection{Velocity Distributions}
\begin{figure}
	\includegraphics[width=\textwidth]{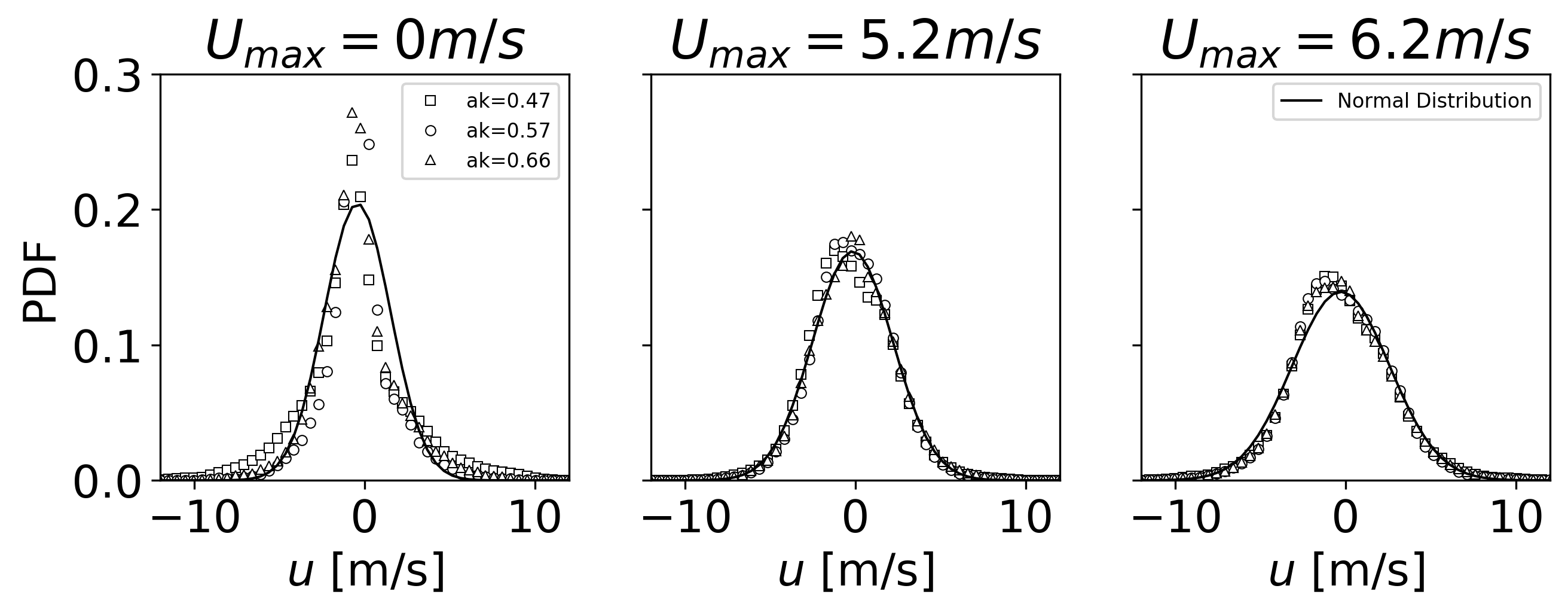}\\
	\includegraphics[width=\textwidth]{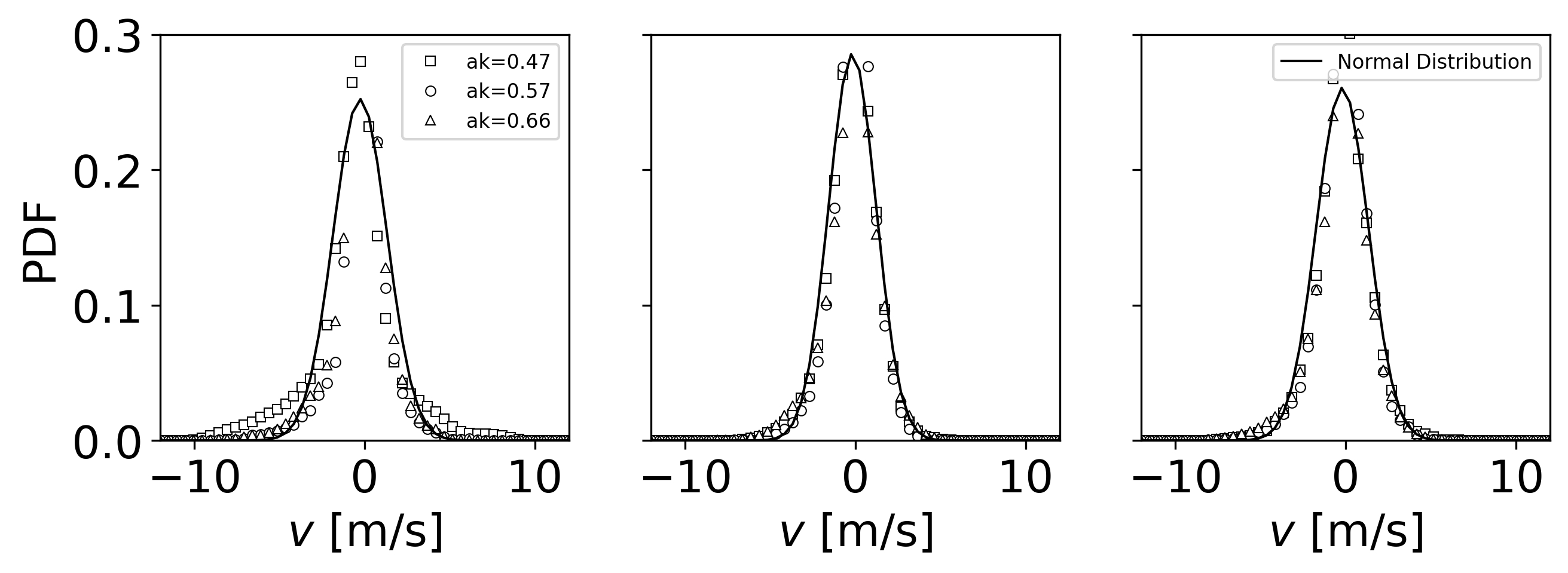}\\
	\includegraphics[width=\textwidth]{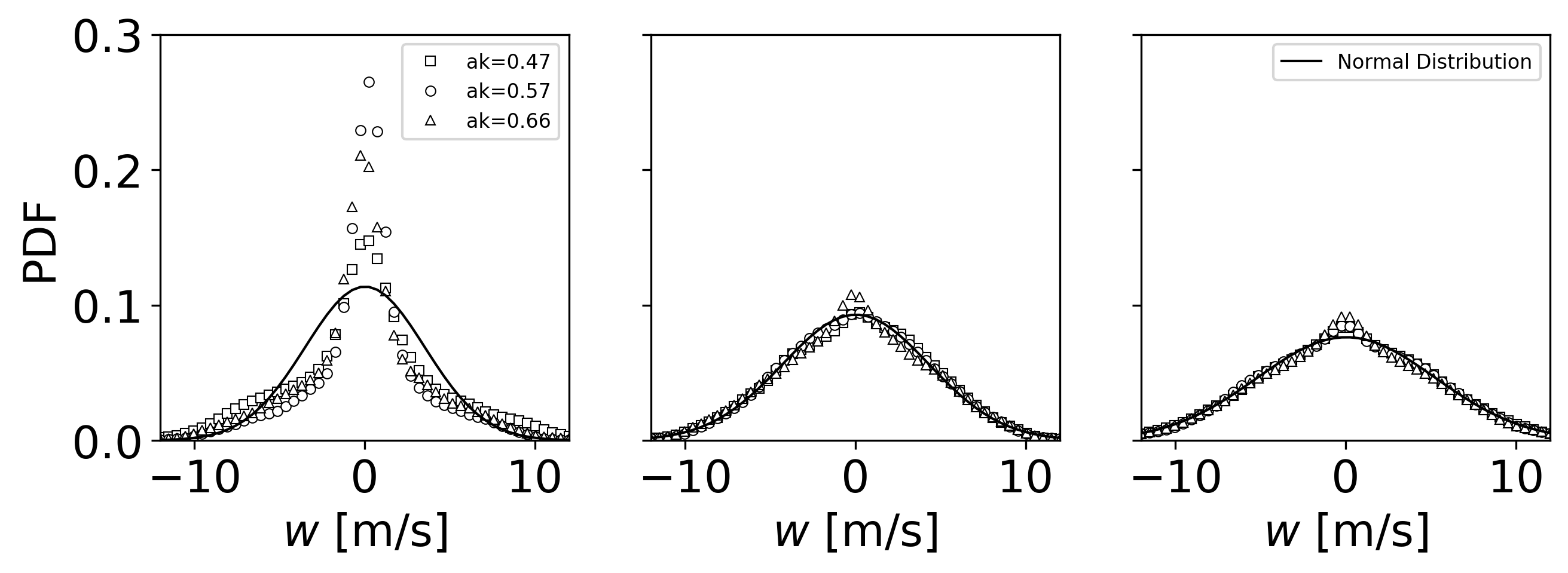}
	\caption{Probability distribution for all the velocity components in the different wind cases. Each vertical panel shows the same $U_{max}$ and each horizontal panel shows one of the velocity components. The different $ak$ are shown with distinct markers and a solid line shows the Gaussian distribution with the same mean and standard deviation as the data. Maximum values of $Re$ are $Re_{0}=120$, $Re_{5.2}=2700$ and $Re_{6.2}=3200$; where the subscript refers to the correspondent $U_{max}$}
	\label{fig:vel-comp}
\end{figure}
\textcolor{black}{The instantaneous velocity components of the droplets have been retrieved by means of PTV. For all cases, the data has been treated as a statistical ensemble and the presented distributions considered all the observed droplets at all time steps during the length of the series. All the data presented has shown convergence of the distribution. The convergence analysis was done by taking 10\% of the data and adding it to the histogram each time.} Figure \ref{fig:vel-comp} shows the probability distributions of the instantaneous droplet velocity components $(u,v,w)$ for all the droplets analyzed in the different cases of $ak$ and $U_{max}$. Only the cases with $U_{max} > 0$ exhibit similarity with the normal distribution. \textcolor{black}{When $U_{max}=0$, the probability for droplets with instantaneous velocity equal to the mean of the distribution is larger than the probability estimated by the normal distribution, especially in the $u$ and $w$ components which refer to the horizontal components. This could mean that some of the droplets do not present horizontal displacement at $U_{max}=0$, but they are affected by the airflow in the presence of wind. On the other hand, the instantaneous vertical component $v$ presents a probability distribution with more extended tails to the extreme values when there is no wind.} This means that the largest vertical velocity is dampened by the presence of wind. 

For all components, the standard deviation increases with $U_{max}$, which is likely an indication of the force applied on the droplets by the wind. The force could be responsible of increasing the variability of the instantaneous velocities, creating larger tails in the probability distribution. In this order, we could consider the flow regime in which the experiments were developed by estimating $Re_{f}$ in the presence of wind. For the case where $U_{max}=0$, $Re_{f}$ is zero, but instead we consider the maximum particle Reynolds number $Re = D_{max}\lvert\Vec{u}\rvert_{max}/\nu$, where $D_{max}$ is the maximum diameter of found droplets and $\lvert\vec{u}\rvert_{max}$ is the maximum speed for the droplets. These values are presented in Figure \ref{fig:vel-comp}. From these values of $Re_{f}$, both wind cases could be consider as turbulent flows. Therefore, droplets with $D_e < \mm{1}$, can follow the flow almost passively. However, it might not be expected to see the effect of turbulent flow onto larger droplets. These droplets will not follow the air flow, nonetheless, there is a change in their behaviour when the air flow is introduced. In addition, the obtained distributions for the velocity components resembles those found for tracer particles in turbulent flows \citep{voth2001measurement,ouellete2006}.

\begin{figure}
	\centering
	\includegraphics[width=\textwidth]{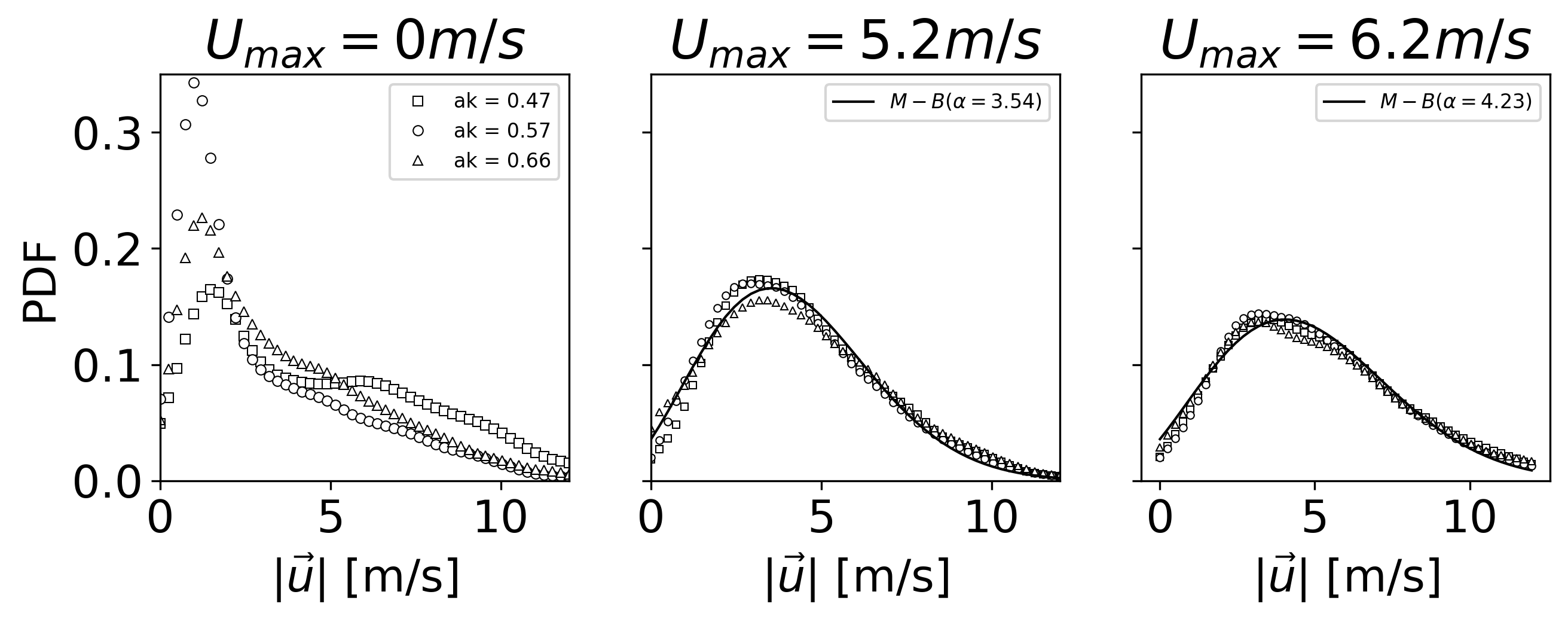}
	\caption{Probability distribution of the droplet speed for the different wind cases. The first graph shows the distribution for $U_{max}=0$, no fit has been added in this case as the distributions differ very much between each other and from the \textit{M-B} distribution. On the other hand, the nest two graphs show the results for $U_{max} > 0$. In these cases, the solid line represents the \textit{M-B} distribution with scale parameter $\beta$, from where we can obtain the mode $\mu = \sqrt{2} \beta$, which is the most frequent value in the distribution.}
	\label{fig:speed}
\end{figure}

From the velocity components, the speed $\lvert\vec{u}\rvert$ can be calculated and the distributions obtained are presented in Figure \ref{fig:speed}. The distributions are compared to the \textit{M-B} distribution, which represents the speed of particles moving in three dimensions with normally distributed velocity components\cite{mandl2008statistical}. In general, it is visible that the speed distributions for $U_{max}=0$ are dependent on the values of $ak$ and differs largely from the \textit{M-B} distribution. These cases present larger probability for extreme values, both towards zero and the maximum speed. On the other hand, cases with wind become independent of $ak$ and follow closely the \textit{M-B} distribution (i.e. the components of the velocity vector have Normal distribution). Moreover, they seem to be dependent on $U_{max}$, the statistical mode of the speed distribution for the wind cases relates to the wind speed as $\mu( u_i ) \sim 0.96 U_{max}$, where the mode represents the value that occurs more often, and for the \textit{M-B} distribution is defined by $\mu = \sqrt{2} \beta$, with $\beta$ the scaling factor of the distribution. \textcolor{black}{This is another indication that the wind speed exerts a force in the droplets and therefore influences the instantaneous velocity of the detected droplets.} Statistically, this is a significant finding, because the normal distribution of the velocity components and \textit{M-B} distribution of the speed have been related to random processes, as in the case of the Brownian motion \citep{batchelor1953theory} or in some cases of turbulent processes. For example, normally distributed velocity components have been found previously in a wind tunnel, downstream of turbulence-generating grid, where the flow is considered to be fully developed \citep{mouri2002probability}. Also, direct numerical simulations of a turbulent flow \citep{vincent1991spatial} has shown similar behavior.

Additionally, we could consider the effect that the wind conditions have on the drag forces of the droplets. Independently of the wind conditions, droplets moving through air will be affected by different factors, specifically the drag force is dependent on the velocity, the size of the droplet and its deformation. Small droplets $D_e < \mm{1}$ will have considerable drag because of the small $Re$, they follow the drag-Reynolds relation for rigid spheres ($C_D = \frac{24}{Re}[1+0.1935Re^{0.6305}]$). \textcolor{black}{For larger droplets ($> \mm{1}$), their deformation is more important, in general, it is known that the drag coefficient is larger for an oblate spheroid than for a sphere, specially when the largest cross-sectional area is perpendicular to the flow.} On top of the drag coefficient variations due to the drop and its own dynamics, the effect of an external flow, like wind, should be consider also. The drag-Reynolds relation becomes fairly complicated because of all the parameters and their variations. Therefore we cannot assume that large droplets will be less affected by drag than small droplets. For example, experimental data of water droplets falling in turbulent flows \citep{laws1941measurements} has shown that for large drops the average drag coefficient is higher than in non turbulent flows. In addition, when studying large particles as flow tracers in Lagrangian methods \citep{xu2008motion}, it was found that the acceleration PDF's where quite similar to those of the tracer particles, with the tails being weakly suppressed. This will suppose that the large particles are to some degree affected by the turbulent flow. Analysis of the acceleration distributions can also be done to compare to this result.

\subsection{Acceleration Distributions}
\begin{figure}
	\includegraphics[width=\textwidth]{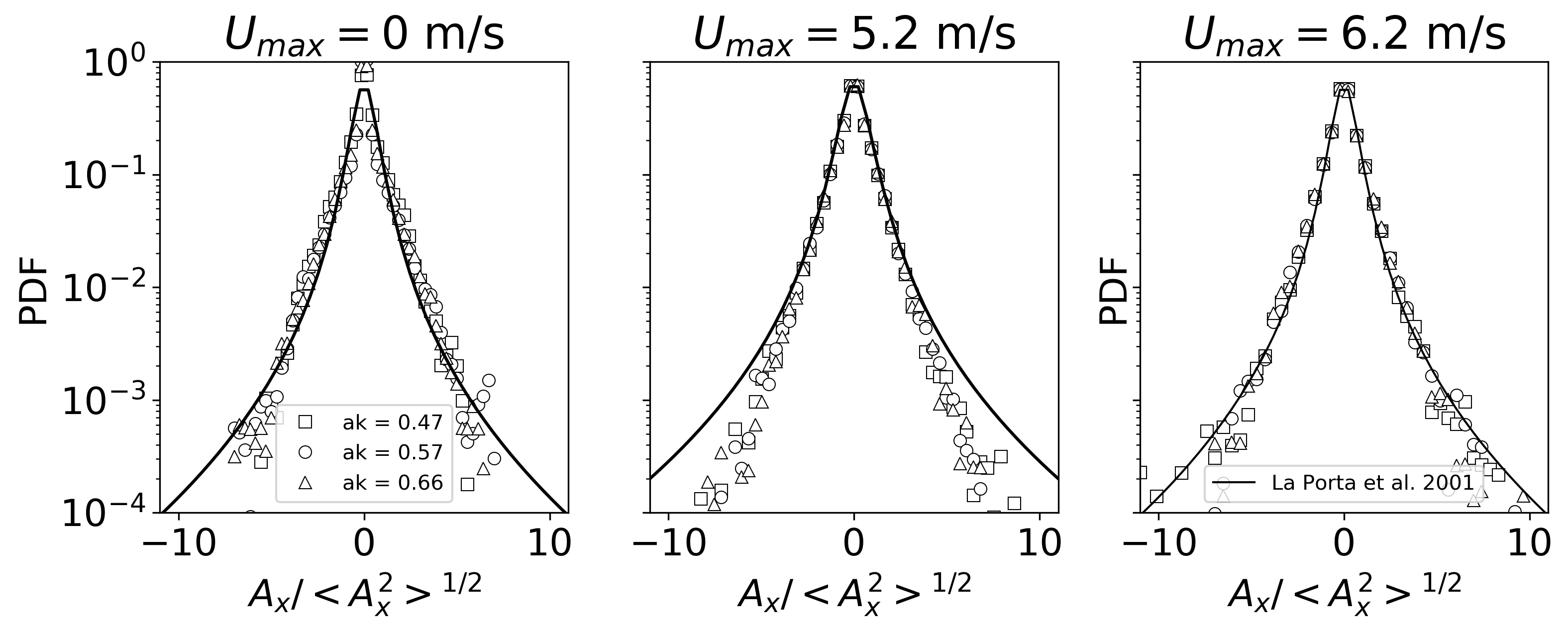}\\
	\includegraphics[width=\textwidth]{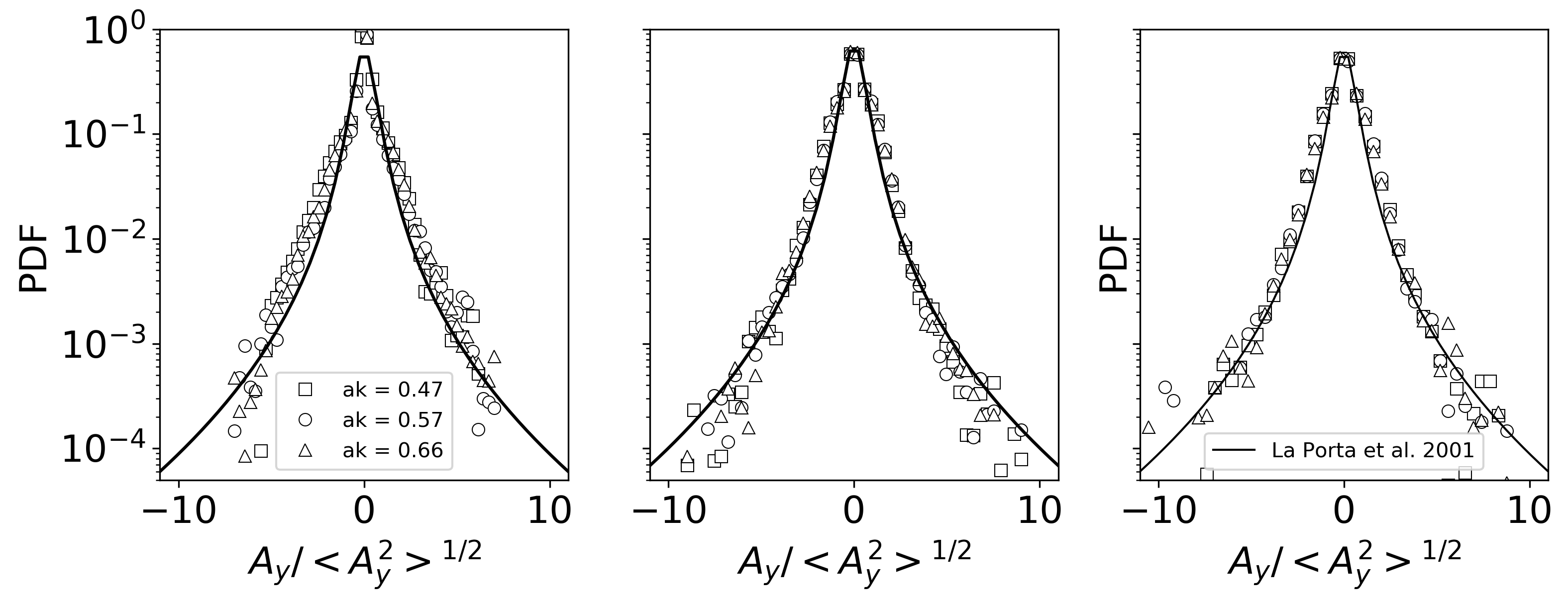}
	\caption{Probability distribution for the acceleration components in $x$ and $y$ for the different wind cases. Maximum values of $Re_{\lambda}$ are $Re_{\lambda,0}=42$, $Re_{\lambda,5.2}=201$ and $Re_{\lambda,6.2}=218$; where the subscript refers to the correspondent $U_{max}$.The different $ak$ are shown with different markers and each vertical panel shows a different cases of $U_{max}$. The solid line shows the proposed exponential distribution \cite{la2001fluid}.\textcolor{black}{Please note that the $\overline{D_e}$ is a function of both $ak$ and $U_max$, as shown in figures \ref{fig:all_waves} and \ref{fig:Pow-diam}.}}
	\label{fig:acc}
\end{figure}
\textcolor{black}{The instantaneous acceleration components of the droplets has been retrieved also by means of PTV. For all cases, the data has been treated as a statistical ensemble and the presented distributions considered all the observed droplets at all time steps during the length of the series.} 
Figure \ref{fig:acc} shows the probability distribution of the instantaneous droplet acceleration components normalized by their standard deviation: $A_{i}/<A_{i}^2>^{1/2}$, $A_x$ in the wind direction and $A_y$ in the vertical direction. The plot also shows the exponential distribution proposed by \cite{la2001fluid} and defined by:
\begin{equation}
	P(A) = C e^{ \left(-\frac{A^2}{(1+\lvert A\beta/\sigma\rvert ^{\gamma})\sigma^2}\right)}
\end{equation}
with $\beta = 0.539$, $\sigma = 0.508$, $\gamma = 1.588$, for the results presented here the constant $C = 0.67$ was obtained from a least squares fit. Only for $U_{max}=\ms{6.2}$ the extremes of the distributions resemble those of the distribution suggested by \cite{la2001fluid}. In the figures, it is visible that the tail values for the experimental data are not as high as those proposed by the distribution. This effect could have two explanations. The first explanation could be the level of turbulence in the flow together with the size of the droplets. The level of turbulence is related to the values of $Re_{\lambda}$, which were calculated to be $Re_\lambda \leq 220$ for all the experiments. These values are in the lower limit of those studied by \cite{la2001fluid}, where $Re_\lambda \geq 200$. Another reason for the differences in the tails of the distribution can be the accuracy of the experimental histogram, as a low number for droplets can be counted for the extreme cases. \textcolor{black}{Together with that, the most of the droplet sizes that are analyzed in this work are too big to follow the flow as passive tracers.} As mentioned before, \cite{xu2008motion} discussed a similar effect on particles of large size compared to tracers, where the tail of the acceleration distributions is weakly suppressed when using large particles. Overall, we can confirm that the dynamics of the droplets produced after the wave splash is affected by the presence of wind even from velocities as low as \ms{5}. The Lagrangian approach allows the study of particle statistics and the statistics of turbulent flows, which is vital for the understanding of dispersion, the study of inertial particles and the development of the statistical models and simulations.

\section{Conclusion}\label{sec:con}
\textcolor{black}{The initial distribution of droplets after a wave breaking event has been studied for droplets between $\mm{0.25} \leq r \leq \mm{5.5}$. The influence of wind on this initial distribution has been addressed by comparing cases of $0 \leq U_{max} \leq \ms{6.2}$.} The analysis shows that the distribution of droplets in all cases is in agreement with the PDF presented in previous studies for ligament-mediated spray formation. A shift of the mean diameter is found and correlated to the energy content of the breaking wave which could point out a relation between the wave energy and the volume of the mean ligament created after breaking. The mean and variance of these distributions are subjected to the properties of the breaker such as breaking type and energy content.

As for the velocities and accelerations, the distributions show noticeable differences between the cases without wind and the cases with wind. The air flow is developed enough to create turbulence, which seems to affect the production and motion of the droplets. When there is wind, the velocity components are normally distributed and the speed follows the \textit{M-B} distribution as predicted by the theory of statistics in turbulent flows. On the other hand, the velocity components differs from the normal distribution when there is no wind, specially the speed has a very distinct shape from the \textit{M-B} distribution and larger probability for extreme values. The findings are similar for the acceleration components, where the distribution for the largest wind speeds has a more extended exponential tail, similar to experimental and numerical studies developed for Lagrangian trajectories in turbulent flows \citep{choi2004lagrangian,gerashchenko2008lagrangian,voth2001measurement,toschi2009lagrangian}. The influence of the flow surrounding the droplets is not negligible for the wind cases as it is shown from the statistics. Furthermore, it can potentially be an important parameter in the droplet phenomenology, such as their vertical reach, their coalescence rates or even their residence times in the atmospheric boundary layers. Further studies should be directed to the understanding of these interactions.

 Recent research showed that the production of large droplets was higher than previously expected \citep{ortiz2016sea}. It is the largest droplets that can more easily breakup and generate more droplets when considering time evolution or increasing wind conditions. Therefore, their presence in the early stages of wave breaking and spray formation needs to be further studied. The presented results gave an insight on the generation of large droplets ($D_e > \mm{1}$) post-wave-breaking. Furthermore, there is evidence of the influence of the flow in the large droplets in this study. Together with recent field studies \citep{lenain2017evidence}, this could suggest that large droplets have a longer lifetime in the atmospheric boundary layer than previously expected. Therefore, the processes from where these larger droplets are created and transported need to be better understood.

\bmhead{Acknowledgments}

Funding from the Norwegian Research Council through the project `Rigspray' (grant number 256435) is gratefully acknowledged. The authors would also like to acknowledge Alex Liberzon and the OpenPTV consortium with their help on the use of the software. The help of Olav Gundersen on the experimental setup and Blandine Feneuil on the revision of the manuscript are also gratefully acknowledged.

\section*{Declarations}

%Some journals require declarations to be submitted in a standardised format. Please check the Instructions for Authors of the journal to which you are submitting to see if you need to complete this section. If yes, your manuscript must contain the following sections under the heading `Declarations':

\begin{itemize}
\item Funding: Funding from the Norwegian Research Council through the project `Rigspray' (grant number 256435) has been used.
\item Competing interests: The authors declare no other competing interests.
\item Availability of data and materials: The data used in this article is already available in Zenodo: \url{https://zenodo.org/record/4277527#.YXanzC8Rr0o}
\end{itemize}

%\noindent
%If any of the sections are not relevant to your manuscript, please include the heading and write `Not applicable' for that section. 

\begin{appendices}

\section{Generation of focusing breaking packets and the effects of shoaling and wind}\label{sec:App}

\subsection{Wind Profiles}\label{sec:A-wind}
The wind profiles, without the influence of mechanically generated waves, were measured using particle image velocimetry (PIV). The center of the field of view (FOV) is 10.75 meters from the wave paddle in the location "PIV FOV", indicated in Figure \ref{fig:ExpSetup}. Two Photron WX100 (2048x2048 pixels) cameras with 50 mm lenses are used, each providing a FOV of approximately 18x18 cm. The cameras were positioned in a vertical arrangement, as indicated in Figure \ref{fig:ExpSetup}. The air phase was seeded with small ($\approx$ 6 $\mu m$) water droplets generated from a high pressure atomizer. The centerplane was illuminated by a 147 mJ ND:YAG double pulsed laser. The cameras were set to acquire image pairs at a rate of 30 fps, and a frame straddling technique was employed to control the effective $\Delta t$ between an image pair used for PIV. Hence, 15 velocity fields were acquired per second and $\Delta t$ was varied between 150 and 350 $\mu s$ depending on the air velocity in the flume. The images (800 per experimental case) were processed in Digiflow by Dalziel Research partners \citep{dalziel2017digiflow}, with a final subwindow size of 80x80 pixels, and 50 \% overlap.

\begin{figure}
	\includegraphics[width=0.7\textwidth]{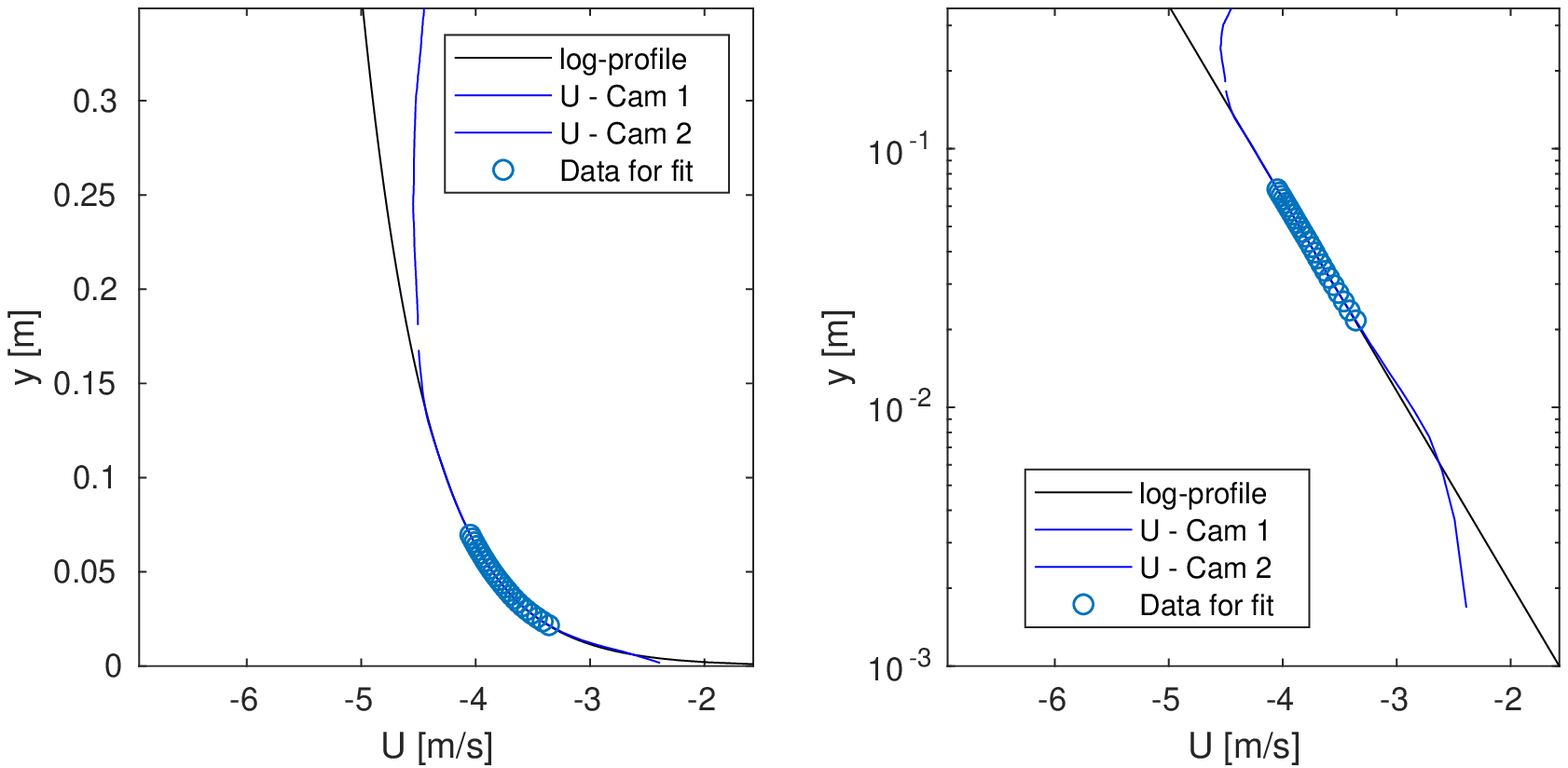}
	\caption{Example recorded velocity profile (blue lines), data points used for fit with equation \ref{eq:logLayer} (blue circles) and resulting log-profile (black line). Illustrated with linear (left) and semilogarithmic (right) axis.}
	\label{fig:LogLayerFit}
\end{figure}

The lower part of the wind velocity profiles were found to be well represented by a logarithmic velocity profile:
\begin{equation}\label{eq:logLayer}
	U = \frac{U_*}{\kappa}\ln(y/y_0),
\end{equation}
where $u_*$ is the wind friction velocity, $\kappa=0.41$ is the Von Karman constant \citep{von1934turbulence} and $y_0$ is the roughness height. Equation \ref{eq:logLayer} was fitted to a part of the velocity profile exhibiting a logarithmic profile, deducing $u_*$ and $y_0$, as shown in Figure \ref{fig:LogLayerFit}. The logarithmic profile was then used to estimate an equivalent $U_{10}$ (mean velocity evaluated 10 meters above the surface). Results are presented in Table \ref{tab:WindVelResultsOverview}, together with the peak horizontal velocity recorded ($U_{max}$). 

\begin{table}[h]
	\centering
	\caption{Results from the wind profile analysis.}
	\begin{tabular}{ c  c  c  c  c }
		\hline
		Wind case & $U_{max}$ [m/s] & $U_*$ [m/s] & $y_0$ [mm] & $U_{10}$ [m/s]\\
		\hline
		1 & 3.41 & 0.151 & 0.0185 & 5.14 \\ \cline{1-5}
		2 & 3.91 & 0.201 & 0.0403 & 6.09 \\ \cline{1-5}
		3 & 5.09 & 0.286 & 0.0984 & 8.03 \\ \cline{1-5}
		4 & 5.45 & 0.308 & 0.1015 & 8.64 \\ \cline{1-5}
		5 & 6.16 & 0.341 & 0.0864 & 9.70 \\ \cline{1-5}
	\end{tabular}
	\label{tab:WindVelResultsOverview}
\end{table}

\subsection{Generation of Focusing Wave Trains}\label{sec:A-wave}
\begin{table}
	\caption{Maximum wave amplitude for the envelope at the focal point $x_f$, for the different voltage inputs in the wave paddle and maximum steepness $ak$ considering all wave trains have $k=7.59$ rad/m}\label{tab:amps}
	\centering
	\begin{tabular}{l c c c}
		\hline
		Wave case  & $a_{max}$ [m] & $ak$ \\
		\hline
		1  & 0.062 & 0.47 \\
		2  & 0.075 & 0.57 \\
		3  & 0.087 & 0.66 \\
		\hline
	\end{tabular}
\end{table}
The mechanically generated waves were created by a horizontal displacement wave paddle, shown in Figure \ref{fig:ExpSetup}. Focusing wave packets were created following the procedure presented in \cite{brown2001experiments}, the focal region is produced by generating waves with increasing periods. To modify the wave energy, different wave amplitudes were generated by varying the maximum voltage input $V_m$, the maximum amplitude $a_{max}$ is shown in Table \ref{tab:amps}. A group of focusing waves is created using this input voltage time history \citep{brown2001experiments}:  
\begin{equation}
	V(t) = b(t) \sin{\Phi(t)}
\end{equation}
for $0\leq t \leq t_s$ with
\begin{equation}
	b(t) = \frac{256}{27}\frac{t^{3}(t_{s}-t)}{t^{4}_{s}}V_{m}
\end{equation}
\begin{equation}
	\Phi(t) = 2\pi f_{0} t \left(1-\alpha \frac{t}{t_{s}}\right)
\end{equation}
where the instantaneous wave frequency is approximately
\begin{equation}\label{eq:omega}
	\omega(t) =\frac{d\Phi}{dt} = 2\pi f_{0} \left(1-2\alpha \frac{t}{t_{s}}\right)
\end{equation}
Under deep water conditions, $\omega(t)$ produces a perfect focus at 
\begin{equation}
	x_{f} = \frac{gt_s}{8\pi \alpha f_{0}}
\end{equation}
therefore, to define $x_f$, the parameters $\alpha$:phase parameter, $t_s$:total time and $f_0$:initial frequency should be constant. In these experiments, the parameters were defined as:
$$\alpha = 0.30,\quad t_s = \s{18}, \quad f_0 = 2 Hz,$$
which defined the focal point at $x_f=\m{11.69}$, approximately the edge of the slope. It is important to notice that the wave number $k$ is only dependent of $\omega(t)$ in eq. \ref{eq:omega}; therefore, using the dispersion relation for intermediate depth: $\omega^2 = g k \tanh(kh)$, values of $k$ can be calculated numerically for each instant in the wave packet. $k=7.59$ rad/m at the breaking point for all cases.

\subsection{Effects of the shoaling}\label{sec:A-shoal}
\begin{figure}
	\includegraphics[width=0.5\textwidth]{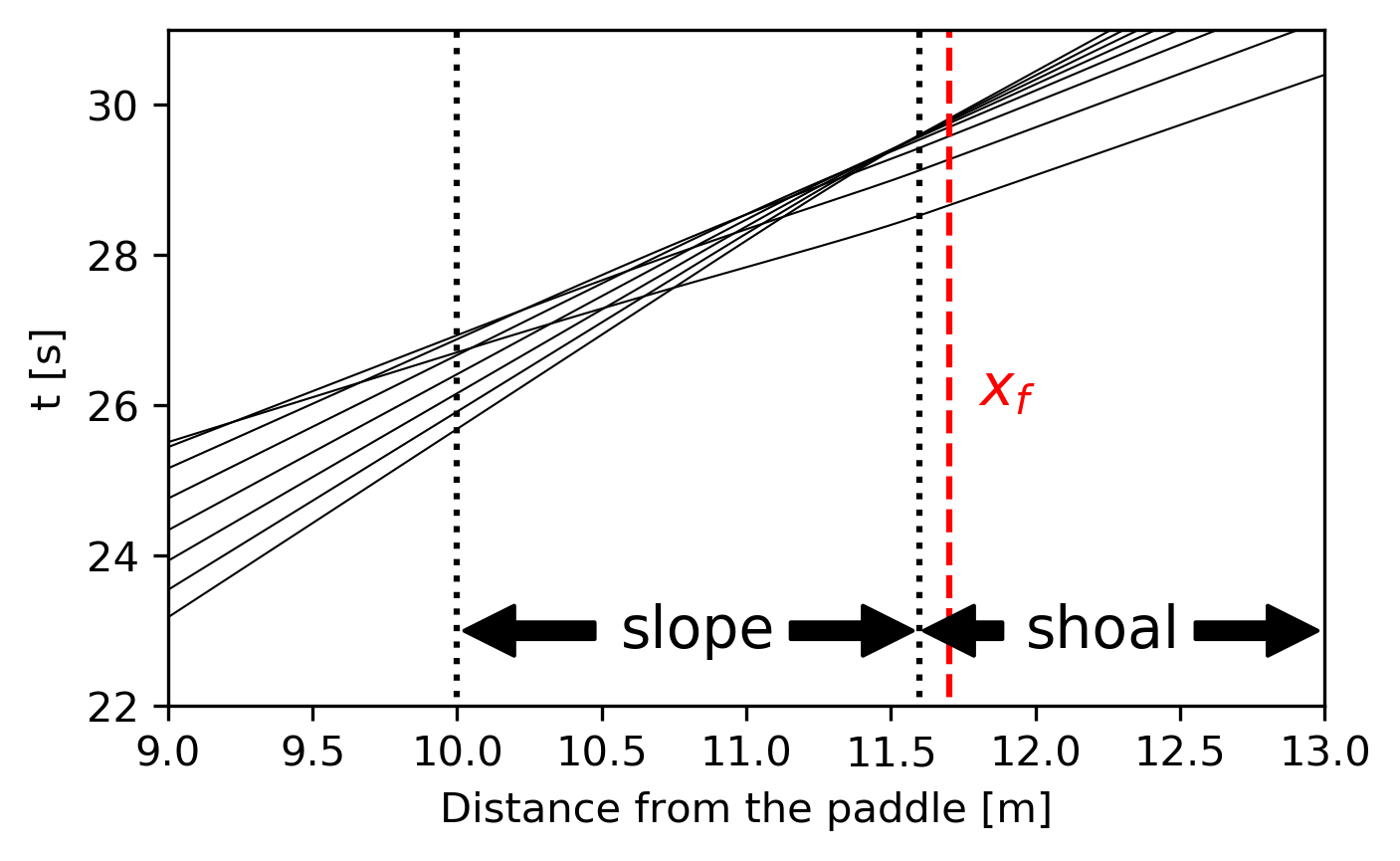}
	\caption{Phase space of focusing wave train, the slope position is limited by the dotted line and the position of the original focusing point $x_f$ is shown with the red line.}
	\label{fig:focus}
\end{figure}

By using this focusing method, the waves can reach steepness $ak > 0.44$ in the focusing point. But the breaking induced by the selected amplitudes only generated spilling breakers and small overturning. It is worth mentioning that larger amplitudes cannot be used, because the wave packets break before reaching the focal point. Therefore, a shoal area was added to steepen the waves even more as they approach the focusing point. In this way, the waves are forced to overturn close to the focusing point as the toe of the wave decelerates and the crest accelerates. The effects of the shoal area over the wave packet at the breaking point have been studied. In Figure \ref{fig:focus}, the phase space of the wave packet is plotted, this shows the effects of the slope at the focal point. The diagram shows that some of the frequencies will reach $x_f$ faster due to the presence of the slope. Nonetheless, most of the frequencies preserve the original $x_f$.

Additionally, the surface elevation at $x_f$ for the wave groups with and without slope can be compared (Fig. \ref{fig:waves2}, to the left). The steepening effect is visible in the second case. The central wave in the packet produces a violent plunger breaker that can be studied. The energy content of the wave group can be quantified by means of the power spectrum. Figure \ref{fig:waves2}(right) shows the power spectrum of the wave groups at $x_f$ for the example case ($ak=0.57$). It is obvious that both cases have the same peak frequency, but the slope case shows evidence of energy dispersion, which was expected. Over all, the presence of the slope affects the energy and spectrum of the wave group but not the position of the breaking point.

\subsection{Wind Generated Waves and their Influence in the Focusing Wave Train}\label{sec:A-ww}
\begin{figure}[h!]
	\includegraphics[width=0.5\textwidth]{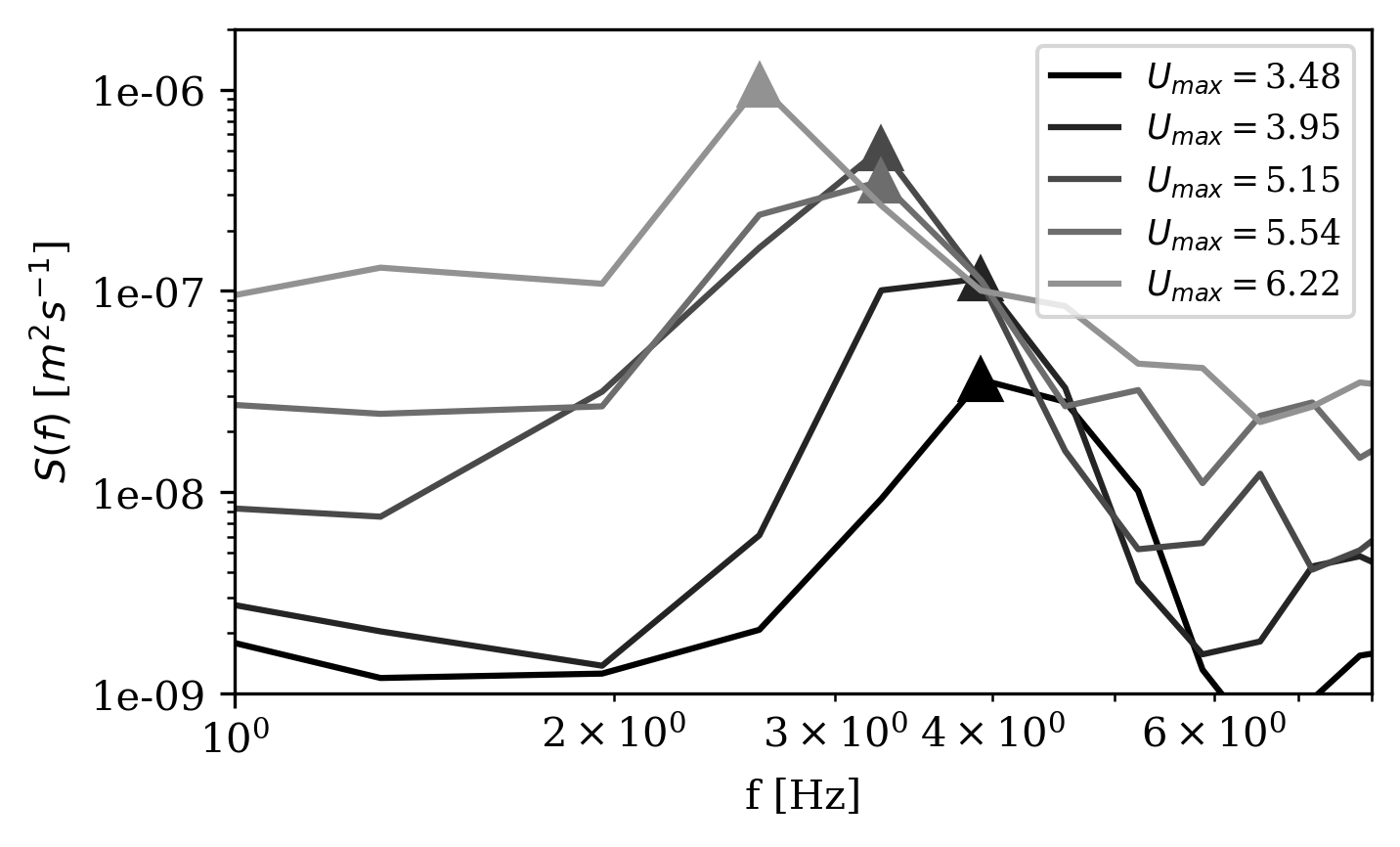}
	\caption{Power spectrum of the wind waves field without the presence of mechanically generated waves.}
	\label{fig:windwaves}
\end{figure}

When introducing wind in the air phase, a field of wind generated waves can appear. Their characteristics will depend on the wind velocity $U_{max}$ and the fetch, as has been studied in wave theory \citep{kinsman1984wind}. The wind wave field will disturb the focusing packet and modify its frequencies and the energy content. Therefore it was important to quantify the influence of this field in the impact zone. Using the wave gauges, one minute time series of the surface elevation were taken for different wind speeds $U_{max}$ without the presence of the focused packet. The 
power spectrum of these series is presented in Figure \ref{fig:windwaves}. The spectra show that the peak frequency of the wind wave field changes with the wind speed. For larger wind speed, the peak frequency decreases and the energy content increases. These frequencies are higher than those for the mechanical generated waves and the magnitude of the coefficients is at most of the same order. From these results, it could be interpreted that the wind could modify the energy content of the wave packet but the influence over the shape and focusing point of the mechanical waves could be minor. To confirm or refute this premise we use the surface elevation and the power spectrum to investigate the wave packet with wind presence.

\begin{figure}
	\includegraphics[width=0.5\textwidth]{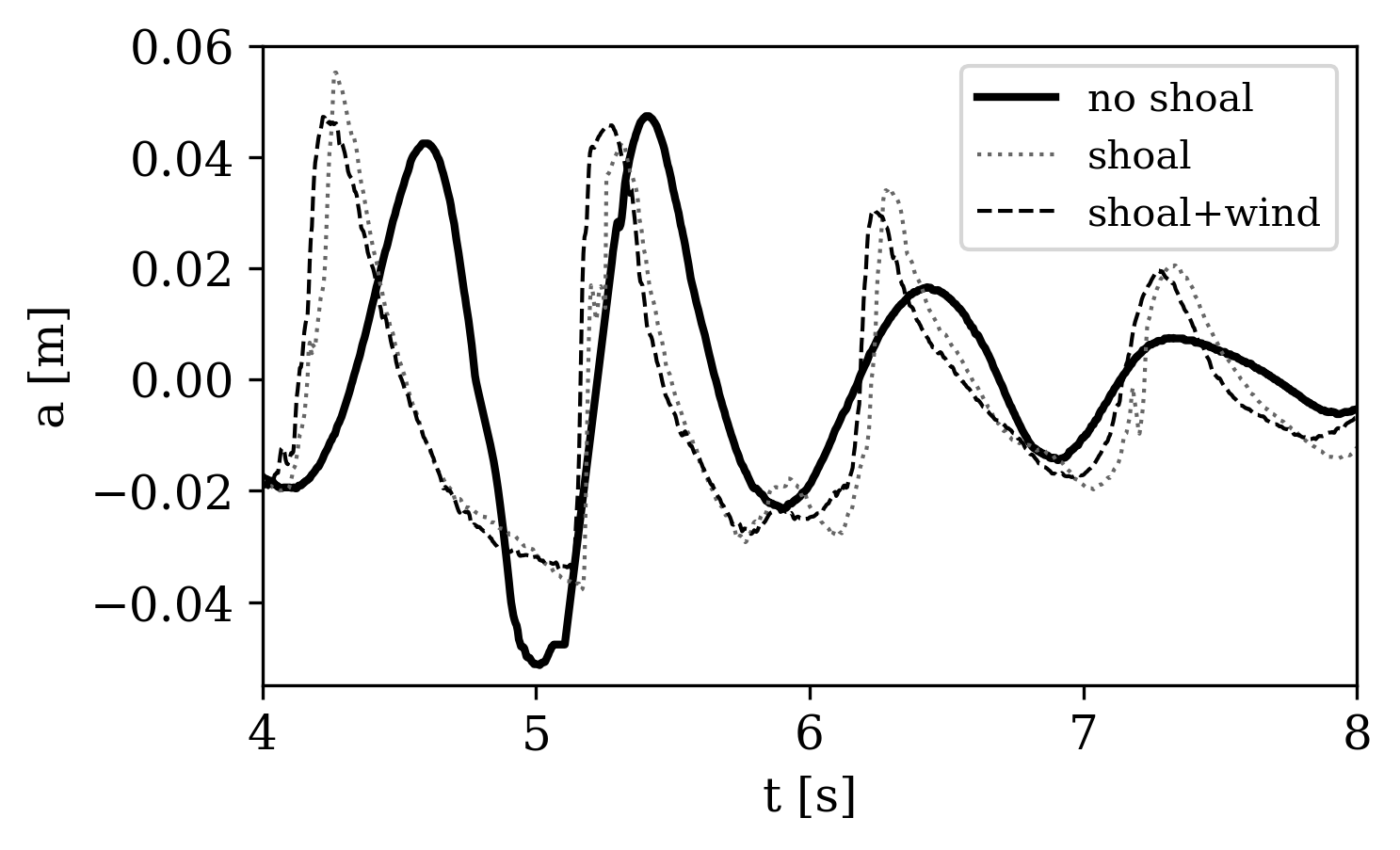}
	\includegraphics[width=0.5\textwidth]{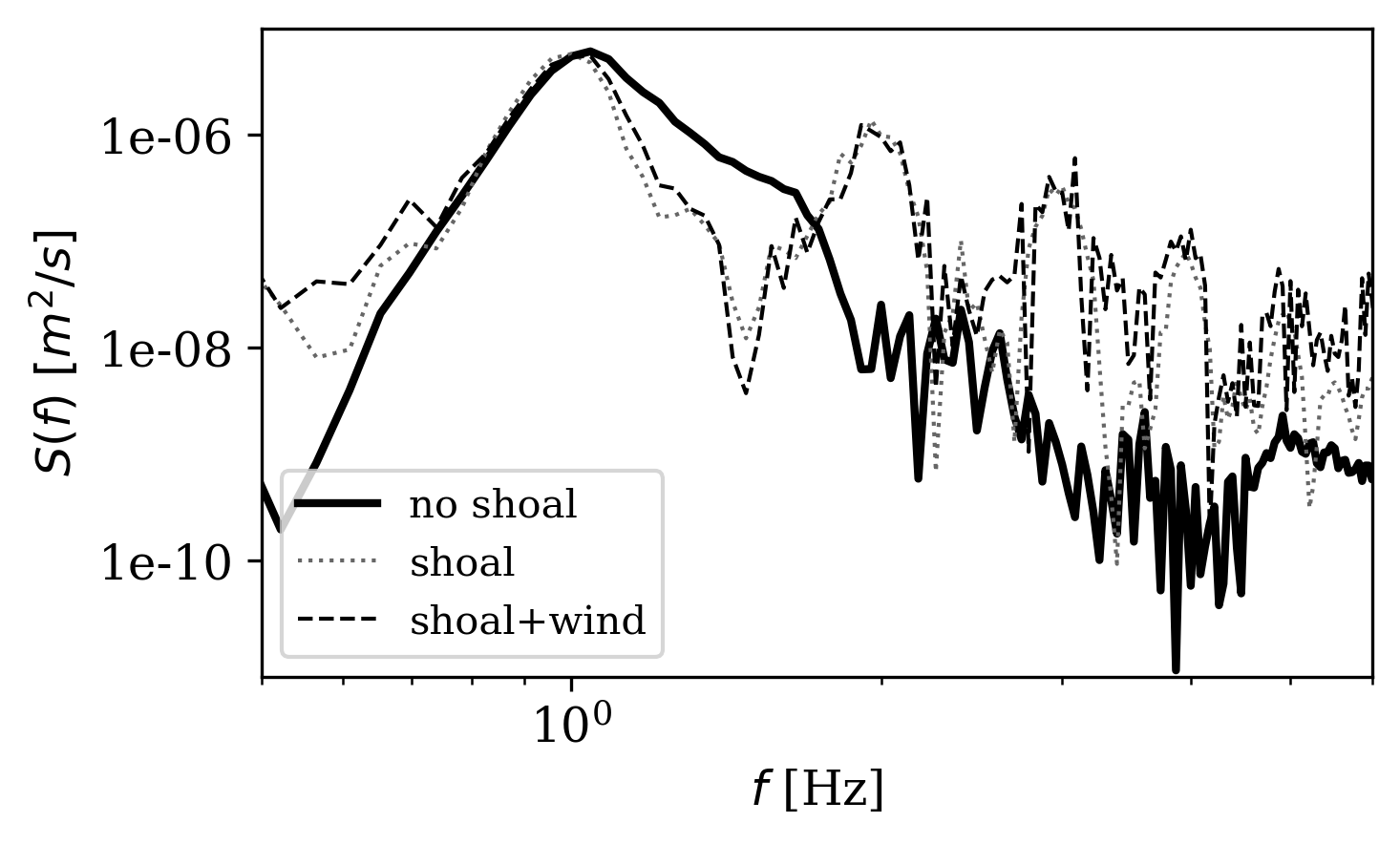}
	\caption{Surface elevation and Power Spectrum at the focal point for cases with wind. Comparison of the same wave amplitude before the slope is situated and after the slope is situated with addition of wind. In this figure $ak=0.57$ has been selected as example for all cases and in the wind case $U_{max}=\ms{6.2}$. The solid line shows the packet previous to the presence of the slope (``no-beach" label in figure) and the dashed line shows the wave packet when the slope is added. With the slope, the wave steepens and overturns to generate a plunger. The presence of wind also affects the steepening but less significantly.}
	\label{fig:waves2}
\end{figure}

In Figure \ref{fig:waves2}, the steepening of the wave from the no-wind case to the wind case is minor compared to the steepening from the slope case to the no-beach case. The frequency domain is also similar in all cases, they have the same peak frequency and the wind cases have dispersion that is indistinguishable from the dispersion created only by effects of the slope.
\end{appendices}


\bibliography{sn-bibliography}% common bib file


\end{document}